\documentclass[12pt]{article}
\usepackage{amsthm, amssymb,amsmath}
\usepackage{multicol}
\usepackage{graphicx}
\DeclareGraphicsExtensions{.jpg,.jpeg,.pdf,.png,.mps,.eps,.ps}
\usepackage{amsfonts}
\usepackage[]{natbib}

\newcommand{\EQL}{\begin{equation}\label}
\newcommand{\EQ}{\begin{equation}}
\newcommand{\EN}{\end{equation}}
\newcommand{\ITM}{\begin{itemize}}
\newcommand{\ITN}{\end{itemize}}
\newcommand{\ENM}{\begin{enumerate}}
\newcommand{\EEN}{\end{enumerate}}
\newcommand{\BEA}{\[\begin{array}}
\newcommand{\EEA}{\end{array}\]}
\newcommand{\EQA}{\begin{equation}\begin{array}}
\newcommand{\ENA}{\end{array}\end{equation}}

\newcommand{\bomega}{\mbox{\boldmath$\omega$}}

\newcommand{\ominfty}{\mbox{$\|\bomega\|_\infty~$}} 

\newcommand{\etall}{{\it et al.}}
\newcommand{\biband}{\&~}
\newcommand{\authone}[2]{#2,~#1}
\newcommand{\authtwo}[4]{#2,~#1,~\&~#4,~#3}
\newcommand{\auththr}[6]{#2,~#1,~#4,~#3,~\&~#6,~#5}
\newcommand{\authfour}[8]{#2,~#1,~#4,~#3,~#6,~#5,~\&~#8,~#7} 
 
\newcommand{\authmanythr}[6]{#2,~#1,~#4,~#3,~#6,~#5,}

\newcommand{\yjfm}[5]{ #1~ #5. {\em J. Fluid Mech. }{\bf #2}, #3-#4.}

\newcommand{\yprl}[5]{ #1~ #5. {\em Phys. Rev. Lett. }{\bf #2}, #3-#4.}

\newcommand{\ypf}[5]{ #1~ #5. {\em Phys. Fluids }{\bf #2}, #3#4.}
\newcommand{\ypfa}[5]{ #1~ #5. {\em Phys. Fluids A} {\bf #2}, #3-#4.}

\newcommand{\yjour}[6]{ #1~ #6. {\em #2} {\bf #3}, #4#5.}
\newcommand{\yproc}[7]{ #1~ #4. In {\em #5} (ed. #6), pp. #2-#3. #7.}

\newcommand{\sjour}[3]{ #1~ #3. {\em #2} (submitted).}

\title{\large Computational Euler History}
\author{Robert M. Kerr \\
Department of Mathematics, University of Warwick}
\begin{document}
\maketitle
\begin{abstract}
A new pseudospectral calculation of collapsing Euler vortices \cite{HouLi06}
has called into question the long-term conclusions of 
singular behavior described earlier in \cite{Kerr93,Kerr05}.  
This review is designed to: improve the discussion  
of the detailed analysis of one test initial condition
designed find sources of errors, to compare
that with calculations showing no evidence of a singularity, and to 
document two sets of discussions. 
Those prior to 1993 between competing teams. And 
recent discussions of what is needed to reach more convincing conclusions.  
\end{abstract}
\section{Introduction}
Whether or not there are singularities of the incompressible three-dimensional
Euler equations is of fundamental mathematical and physical importance. This
review will not discuss either of these issues or related questions 
for how this behavior could be applied to our understanding of 
partial differential equations or turbulence theory.  Instead, the focus
will be the history, circa 1990, of 
numerical calculations that were designed to determine if there could be 
a singularity of the three-dimensional incompressible Euler equations 
and how, through several workshops, a concensus was reached about to address 
the issue numerically.  Analytic results will not be discussed other than how 
they pertain to these calculations.  One test
calculation from \cite{Kerr92} will be used to suggest that 
the differences between the structures in the new calculation
by \cite{HouLi06} and those presented by \cite{Kerr93} could be
associated with small-scale noise.  How this could disrupt 
a singular structure forming is discussed.

The paper is organized as follows.  After the early history of 
Euler calculations is given, methodologies agreed upon by concensus
at several workshops are presented, including how to compare calculations.
Next is a discussion of the uses and dangers of the most robust approach,
namely pseudospectral methods. Then a section compares \cite{Kerr93} 
to competitive
calculations from the early 1990s.  Finally there is a comparison between
\cite{Kerr93} and \cite{HouLi06}.

\section{Early Euler calculations}

The first serious attempt to study Euler numerically 
was the Taylor-Green calculations of \cite{Brachetetal83}.
Their conclusion was that vortex sheets suppressed any
trends towards singularities that had previously been
suggested either by spectral closures or 
series expansions.  This conclusion was qualitatively similar to the conclusion
from two-dimensional ideal, incompressible MHD calculations (\cite{Sulemetal85})
where it was found that current sheets suppressed trends towards singularities.

The next set of numerical investigations used vortex filaments.  While these are
a poor approximation of the Euler equations at the smallest scales, 
they did provide a global picture that an anti-parallel initial condition 
might predispose a calculation towards singular behavior in a manner that 
smooth initial conditions might not (\cite{PumirSiggia87}).  
This led to two preliminary studies at low resolution 
(\cite{AshurstMeiron87,PumirKerr87}) whose only significant contribution was 
that vortex flattening might play a role
similar to current sheets in 2D MHD.  Simultaneously, 
an orthogonal initial condition was simulated that showed 
more curvature development than the anti-parallel
simulations (preliminary results by Melander and Zabusky in 1988 eventually
appeared in \cite{BPZ92}).

All the subsequent calculations assumed an anti-parallel geometry, for which
there are two symmetry planes.  One in $y-z$ is between the vortices and was
called the `dividing plane'. The other in $x-z$ is at the position of 
maximum perturbation and was called the `symmetry plane'. (Terminology
due to F. Hussain.)
The next step was to find a better initial condition within this geometry
and the first steps towards adaptivity.  \cite{MelanderH89}
were able to identify an initial condition
where all derivatives in the profile of the vortex core went smoothly
to zero and used a sinusoidal perturbation of the $x-z$ position as
a function of $y$.  The calculation did not take advantage of the
symmetries, had uniform mesh spacing in all direction and was viscous.
\cite{KerrH89} tried a truncated Gaussian core and
resolved the discontinuities at the edge of the initial condition
by applying a strong Fourier based hyper-viscous filter.  
\cite{KerrH89} used a combination of higher order trigonometric functions 
to create an initial perturbation that was localized
near the symmetry plane.  Symmetries were used to reduce the computational
requirements, the mesh spacing depended upon direction and the calculations
were viscous.  \cite{PumirSiggia90}
uses a hyperbolic trajectory for the vortex lines, 
therefore being the only calculation to
date that is not in a periodic geometry.  
They used a strictly Gaussian core that does not ever go to zero, which
is possible if one does not have to worry about overlapping across
periodic boundaries.  Their calculation was completely adaptive and inviscid.

Of these calculations, only the viscous calculation of \cite{KerrH89}
had any consistency with singular behavior.  To test the \cite{MelanderH89}
profile in adaptive calculations, \cite{Kerr93} and \cite{ShelleyMO93} 
tried Chebyshev and mapped Fourier methods respectively.
\cite{Kerr93} reported consistency with the singular trends 
of \cite{KerrH89}.  
\cite{ShelleyMO93} concluded that the trends did not
favor a singularity, but their calculations used only the sinusoidal 
perturbation of \cite{MelanderH89}
and were viscous. Why they reached different conclusions 
was discussed briefly by \cite{Kerr92}
and will be discussed further here in subsection \ref{sec:Shelley}.

The only significant calculations
since then are due to \cite{Graueretal98} and now \cite{HouLi06}.
The \cite{Graueretal98} calculation seems consistent with the trends
in \cite{Kerr93} while \cite{HouLi06} is not.  
\cite{Kerr93} and \cite{HouLi06} 
are compared in section \ref{sec:HouLi}.  The Kida vortex promoted
by \cite{Pelz01} as a scenario for a singularity has recently
been shown to be regular by Grauer (private communication).

\section{Workshop methodologies}

The methodology of \cite{PumirSiggia90} and 
\cite{Kerr93} was worked out at two workshops on 
Topological Fluid Dynamics chaired by H. K. Moffatt in 1989 and 1991.  
These were the IUTAM Symposium on Topological Fluid Dynamics in 
August 1989 in Cambridge, England and the Program on Topological Fluid Dynamics
at the Institute for Theoretical Physics in Santa Barbara in the Fall of 1991,  
with a symposium on issues in Euler at the end of October 1991.

The following principles were worked out at those programs in discussions 
between most of the principal authors at that time such as 
U. Frisch, F. Hussain, R.M. Kerr, R. Pelz, A. Pumir E.D. Siggia, N. Zabusky, 
and the other participants in those workshops.

These are some of the rules:
\ENM \item Run only Euler.  Do not try to reach conclusions about Euler using
a series of decreasing viscosity Navier-Stokes calculations.  
The range of Reynolds numbers available to Navier-Stokes calculations
at that time is too short to be useful in reaching any definite conclusions.
\item Use refined meshes.  Even the easy solution of different mesh sizes 
in different directions, as in \cite{KerrH89}, has limitations.
There is too much space over which nothing is happening.
Complementary pseudo-spectral calculations can still be useful to confirm the
numerical method, but serious compromises have to be made as discussed below.
\item Have a localized initial perturbation.  Orthogonal vortices
and the hyperbolic trajectory of \cite{PumirSiggia90} automatically
satisfy this.  \cite{KerrH89} and \cite{Kerr93} satisfy this condition
using a map of a sinusoidal perturbation.
\item One needs to demonstrate structures that wouldn't be subject to 
depletion of nonlinearity.  In particular if vortex sheets appear, 
they must develop strong curvature or kinks.
\item The primary quantity of interest is the $L_\infty$ norm of 
vorticity $\ominfty$ so that consistency with the analytic result of 
\cite{BKM84} can be tested.  
If singular behavior is suspected, then the following must be obeyed:
\EQL{eq:BKM} \int_0^t \ominfty dt \rightarrow \infty \EN
\item If it has power-law behavior of the form
$\ominfty\sim (T-t)^{-\gamma}$ and $\gamma<1$, 
then the calculation is incorrect.
\item The best way to test \eqref{eq:BKM} is to assume $\gamma=1$.  
This would be dimensionally consistent.
\item However, using \ominfty by itself can lead to misleading conclusions 
either way.  Double exponentials are indistinguishable from power laws 
and at late times it is impossible to
avoid some slowing of singular trends, which could lead to the appearance of
saturation if one assumes that $\alpha=d\ominfty/dt$.
\item Therefore one must calculate the time derivative of \ominfty directly, 
that is $\alpha=\omega_i e_{ij} \omega_j/|\omega|^2$ directly, which
should also have time-integral singular growth (\cite{Ponce85}).  
In the anti-parallel case $\alpha$ is just the velocity derivative 
on the symmetry plane of the axial velocity.
This gives an independent measure of singular activity 
While finding $\alpha_p$ at the point of $\ominfty$ is most desirable, 
finding $\alpha_s=\sup(\alpha)$ in the symmetry plane was found 
to work better in \cite{Kerr93}.  
\item One needs a measure of collapse to satisfy the condition of
\cite{Caffarelleetal82}.
\EEN

Some of these rules were refined by further conversations between R.M. Kerr
and A. Majda during a two week workshop at the Research Institute 
in Mathematical Sciences in Kyoto, Japan in October 1992.  

Besides the rules just listed, other tests that have been tried and their
order of success are as follows:

\ITM \item[k)] Finding independently that 
\EQL{eq:ompr} \Omega_{pr}=\int dV \omega_i e_{ij} \omega_j
\sim \frac{1}{T-t} \EN
with the same singular time as for \ominfty and $\alpha$.
The evidence from \cite{Kerr93} is that this test works better than
finding either $\alpha_p$ or $\alpha_s$ defined above.
\item[l)] \cite{PumirSiggia90} discusses how the position of \ominfty
moves.  \cite{Kerr93} goes further and tries to show
that the positions $(x,z)_p$ of all minima and maxima of components 
of the stress tensor in the symmetry plane collapse to
the positions of \ominfty. That is
$$ x_p-X(T)\sim T-t \quad,\quad  z_p\sim T-t $$
\item[m)] \cite{Kerr05} goes further still by looking at profiles.
\item[n)] $\sup(|v|^2) \sim T-t $ where $v$ is the axial velocity in the 
direction of vorticity in the symmetry plane. 
This would provide consistency with a suggestion from \cite{ConstFM96}. 
The only evidence for this is marginal from
\cite{Kerr05}. We need a good calculation to test this.
\item[{\cal o})] Curvature blowup as $\kappa^{-2}\sim (T-t)$.  
Again, this would provide consistency with a suggestion from \cite{ConstFM96}. 
So far this is only inferred by
related scaling properties in \cite{Kerr05}.   
Again, a good calculation is needed to test this.
\ITN

\section{Uses and pitfalls for pseudospectral \label{sec:uses}}

For any problem where most of the domain is not involved in the dynamics,
as for this possibly singular equation, uniform mesh methods waste most
of the computational domain.  Pseudospectral codes fall into this class
of methods.
Besides being limited in the local resolution they can provide, pseudospectral
codes have the additional annoyance of a continuing debate over how
to handle the high wavenumber cutoff.

Many strategies have been tried over the years.  Initially Orszag and Patterson
decided upon a spherical truncation which was corrected by adding
together two phase-shifted versions of the nonlinear terms.  Then NASA Ames
proposed the Leonard filter for LES calculations.  
By the mid-1980s it was generally
agreed that none of these tricks gained one anything.  LES calculations started
using only a sharp-wavenumber cutoff with the 2/3rds rule 
and DNS in wall-bounded
flows at NASA Ames/CTR did the same.  All of my calculations after 1985,
whether Euler, Navier-Stokes, or convection have followed this policy.

The issue has been reopened by \cite{HouLi06}
claiming to have found
a filter with 36th-order accuracy.  So let me summarize my understanding of 
the good and bad points of different methods for the Euler problem.

Options for high wavenumber cutoff of Euler:
\ITM \item 2/3rds dealiasing.  
\ITM \item Good: absolutely no wavenumber contributions
for $k>(2/3)k_{max}$, which would immediately give errors.  
\item Good: energy is exactly conserved. 
\item Bad: An abrupt
cutoff leads to Gibbs phenomena (oscillations) in physical space.  
\item Solution:
Must set a strict standard when the calculations must be shut off based upon
anomalous growth of the wavenumbers near the cutoff.  This is good in that
there is a measure of the errors.
\ITN
\item A smooth filter that ends for $k=(2/3)k_{max}$.  
\ITM \item Good: minimize Gibbs phenomena.  
\item Bad: Dissipative.  
\item Bad: No measure of errors.
\ITN
\item A smooth filter that extends beyond $k=(2/3)k_{max}$.  
\ITM \item Good: Maybe minimizes Gibbs phenomena.  
\item Bad: Immediately adds aliasing errors. 
\item Bad: might add small regions of negative vorticity that can blow up. 
\item Bad: No measure of errors. 
\item Unsolved: What of the claim in \cite{HouLi06} 
that this converges? 
And how does one identify what the aliasing errors turn into?
\ITN
\ITN

Despite these problems, it was concluded that
in the early stages a pseudo-spectral code 
is much better for testing initial conditions, 
in particular testing schemes for filtering the initial condition.
It was in ensuring that the nonlinearity was not depleted
even at low resolution and early times in such test calculations
that led to the initial
condition and numerics that worked so well in \cite{Kerr93}.

\begin{figure}
\includegraphics[scale=.9]{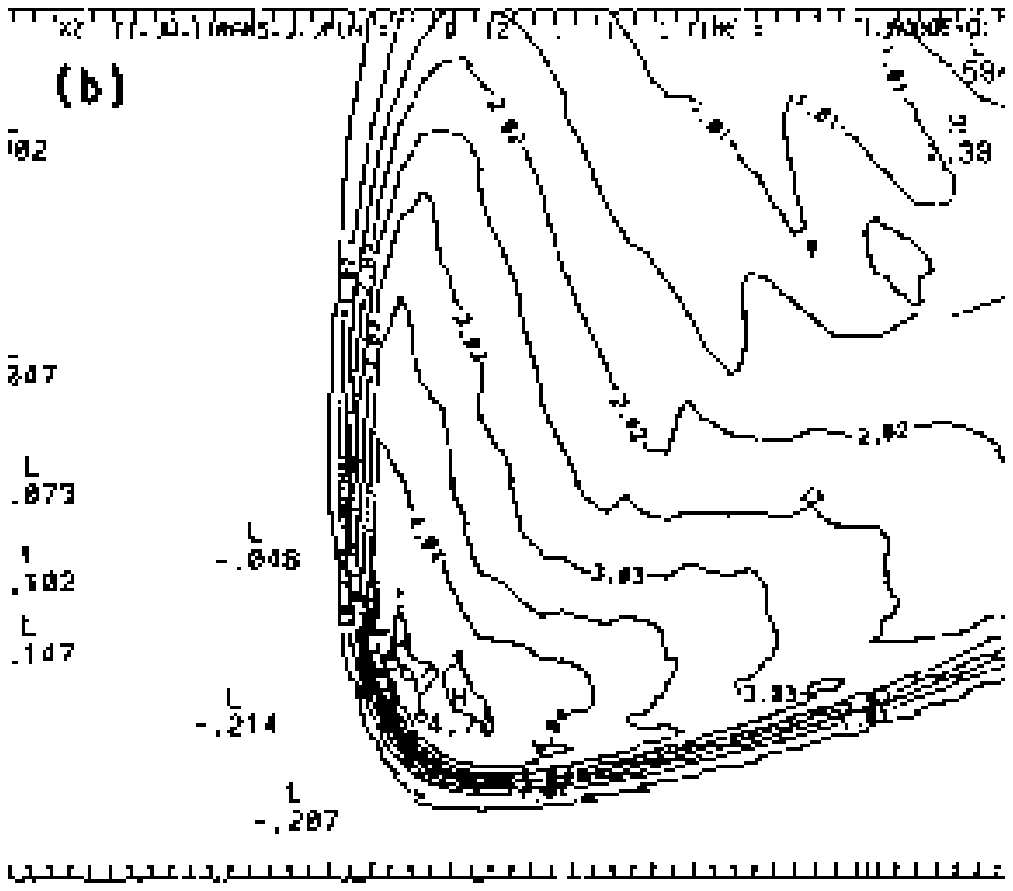}
\includegraphics[scale=.9]{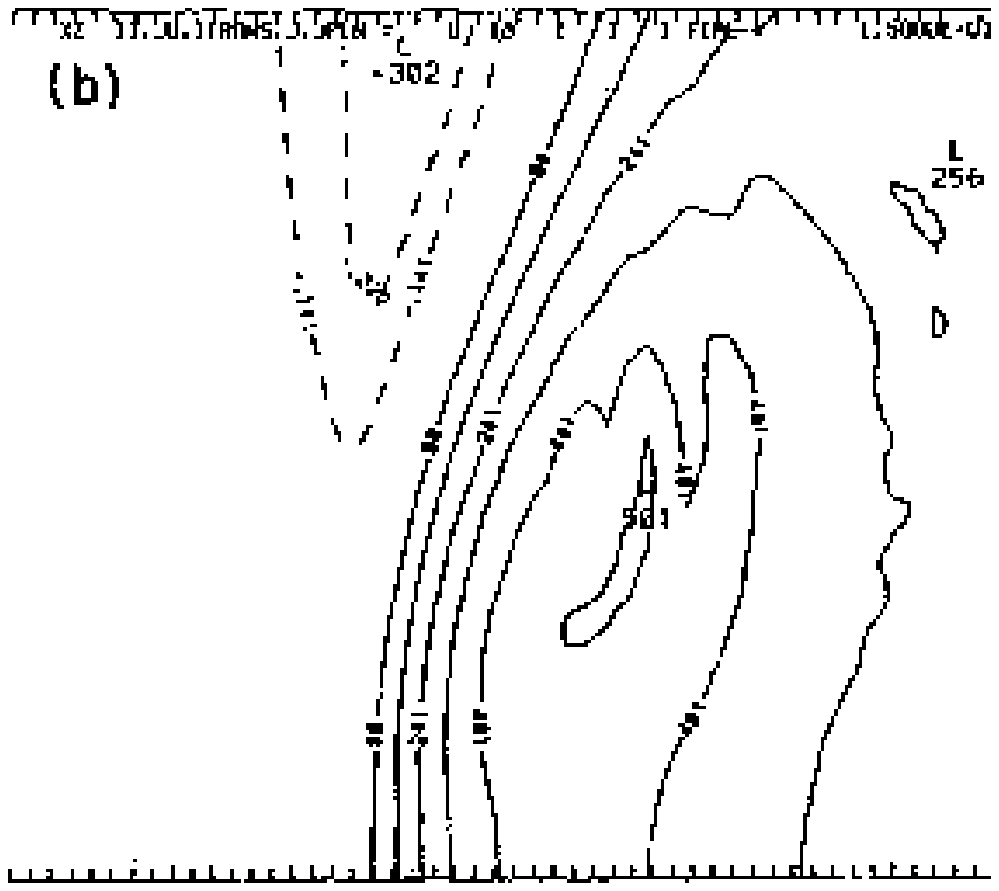}
\caption{$\ominfty$ and $\alpha$ contours in the symmetry plane for $t=15$ from
\cite{Kerr93}}\label{fig:k93t15} 
\includegraphics[scale=0.75]{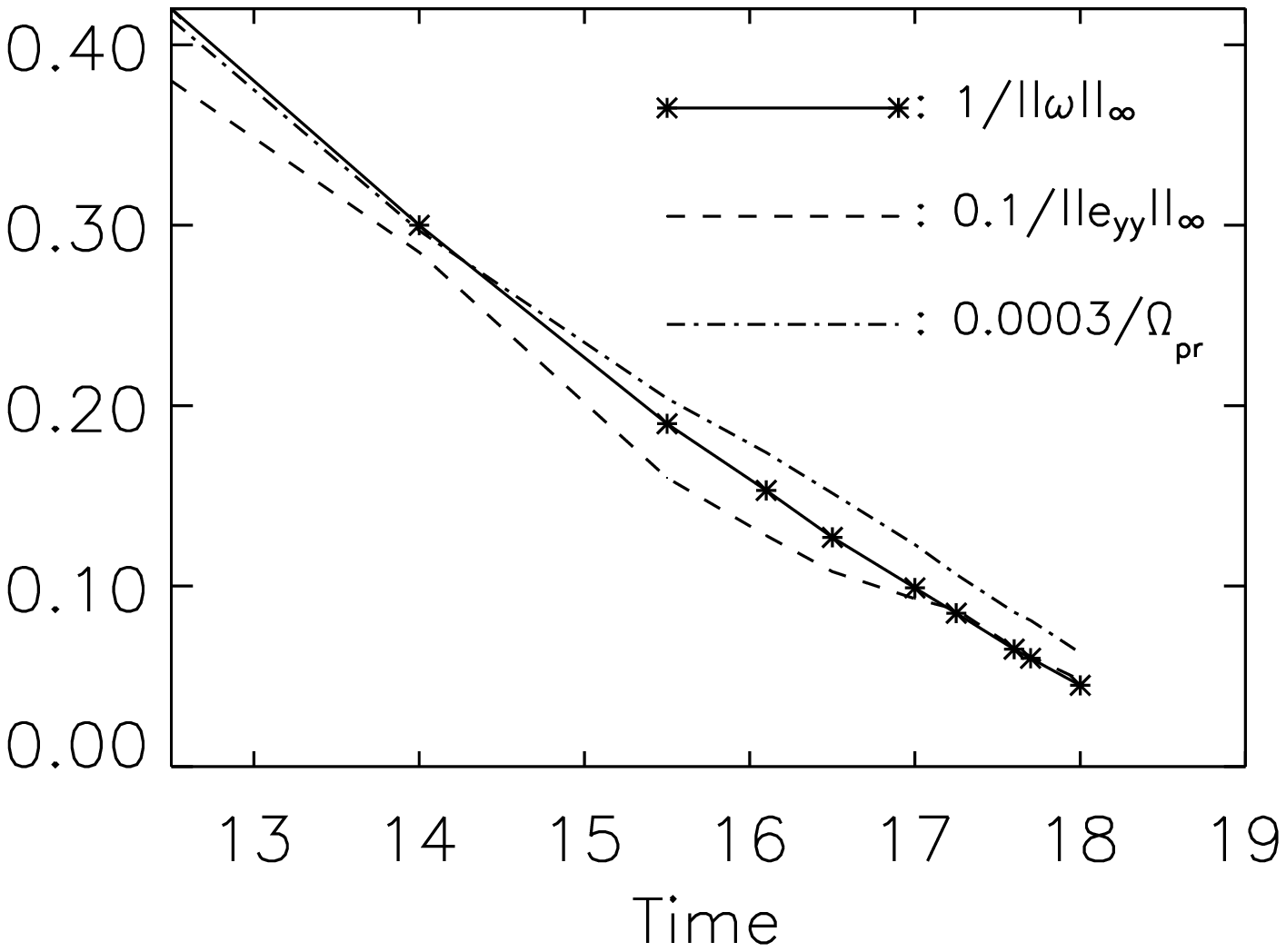}    
\caption[]{Dependence of $1/\ominfty$,
$0.1/|e_{yy,p}|$ and $0.0003/\Omega_{pr}$
upon time from the anti-parallel Euler calculation
showing convergence to a singular time of about $T=18.7$.}
\label{fig:omp}
\end{figure}
\begin{figure}
\begin{minipage}[c]{.5 \textwidth}
\vspace{-5mm}
\includegraphics[scale=0.35]{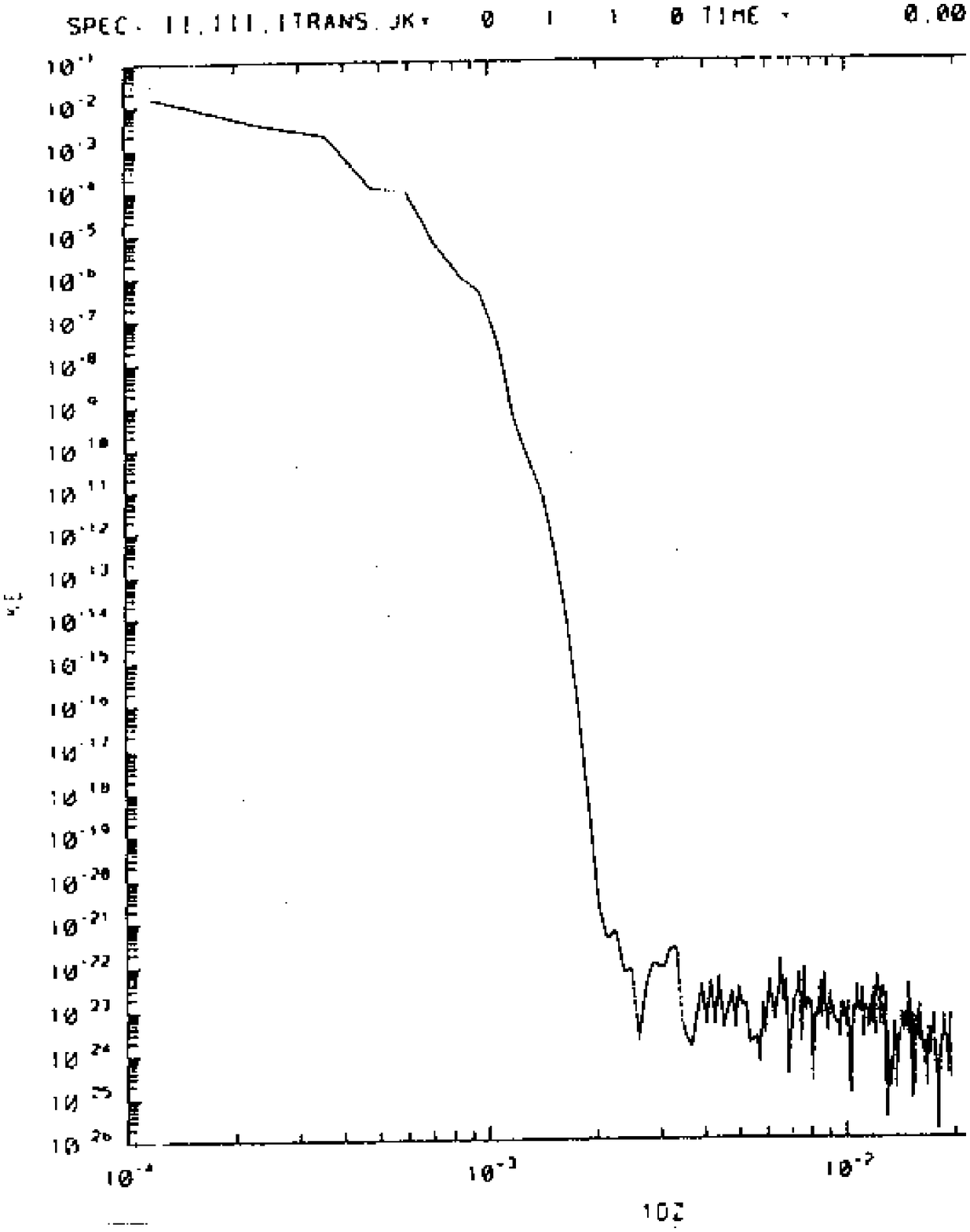}    
\vspace{8mm}
\caption{Initial $z$-Chebyshev distribution (spectrum) for
filtered intial conditions.}
\label{fig:Chebspec}
\end{minipage}
\vspace{10mm}
\begin{minipage}[c]{.5 \textwidth}
\vspace{0mm}\hspace{-8mm}
\includegraphics[scale=0.75]{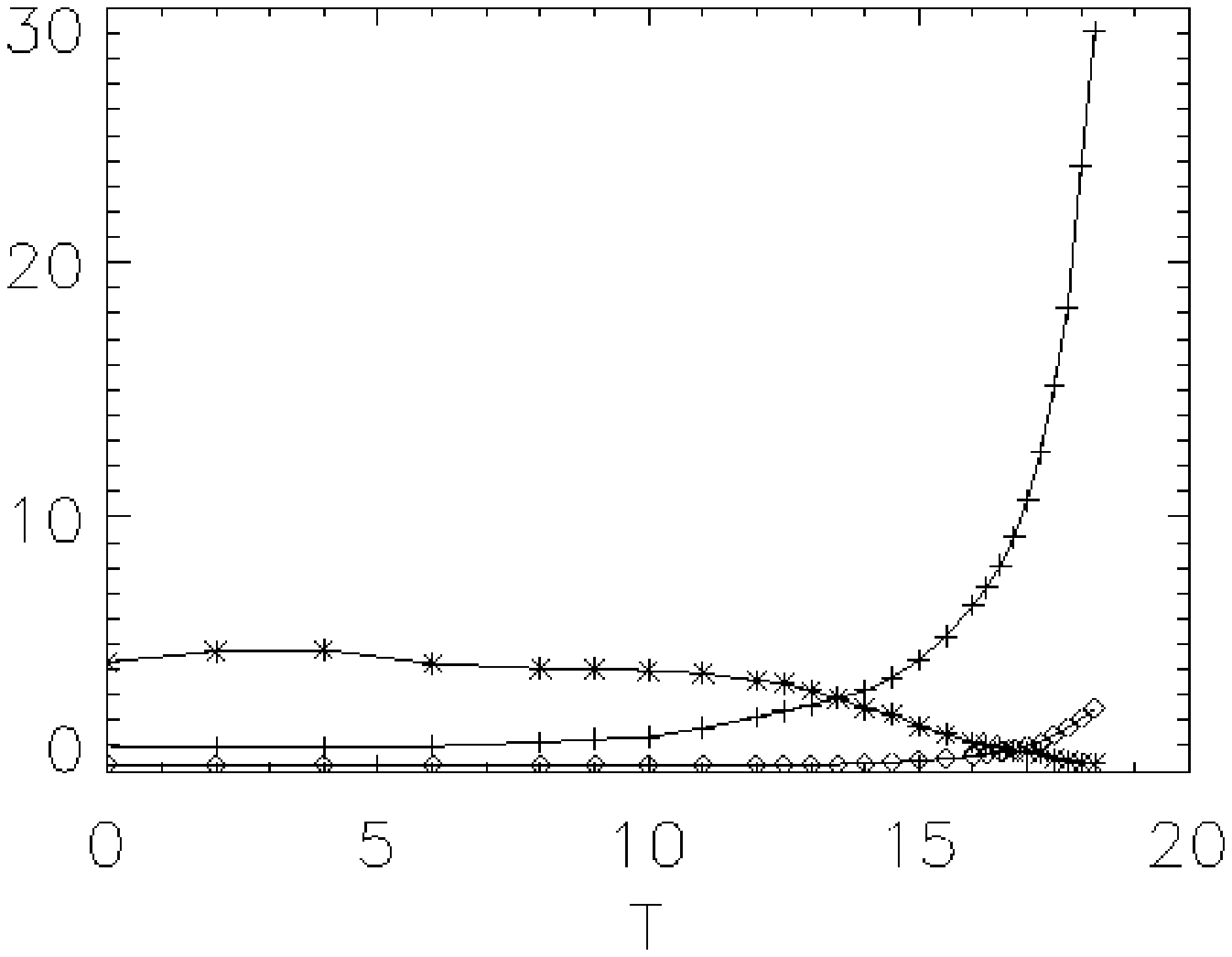}    
\vspace{5mm}
\caption[]{Dependence of $\ominfty$,
$|e_{yy,p}|^{-1}$ and $|e_{yy,p}|$ 
upon time from the filtered initial condition anti-parallel Euler calculation}
\label{fig:omp1}
\end{minipage}
\end{figure}
\begin{figure}
\includegraphics[scale=.75]{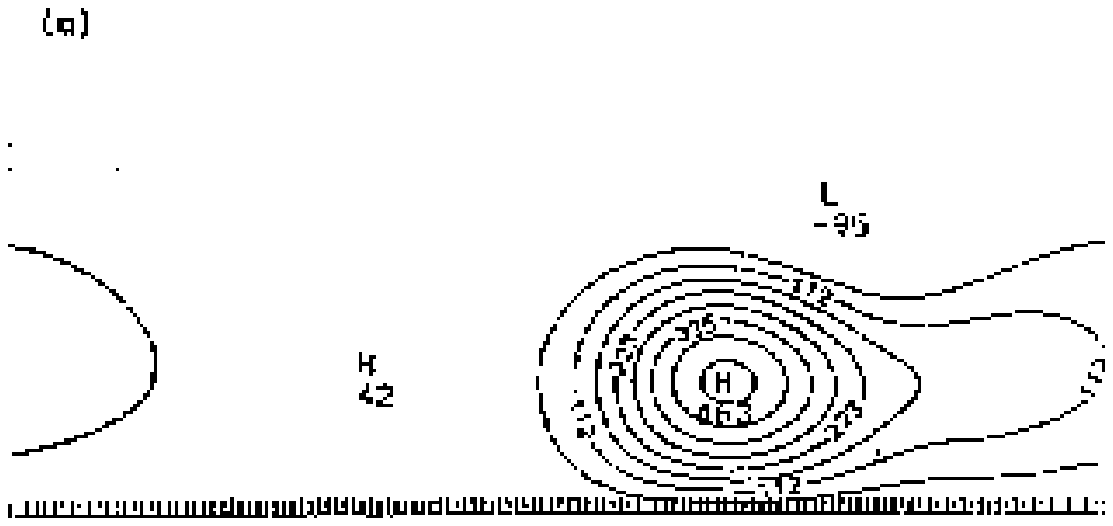}
\includegraphics[scale=.75]{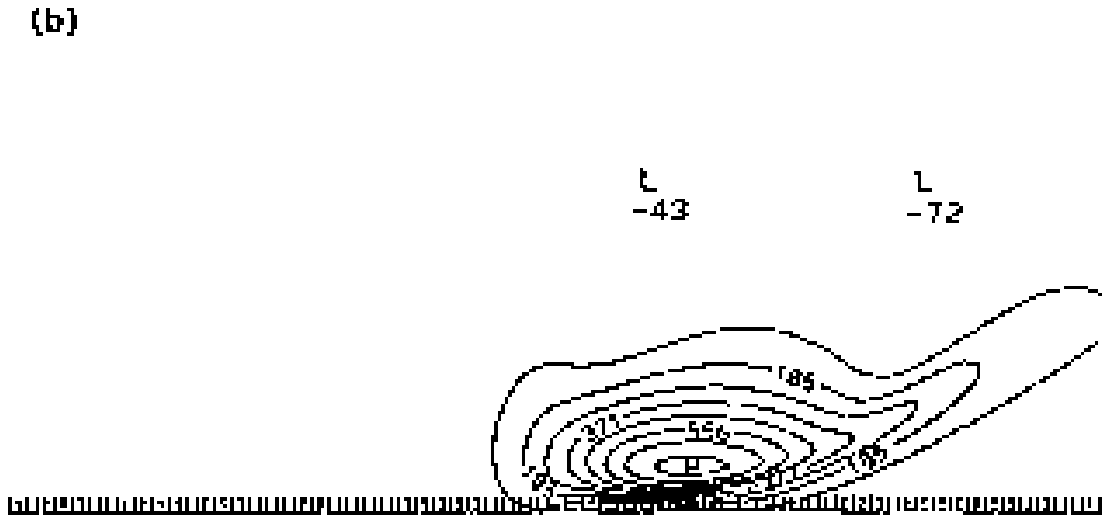}

\includegraphics[scale=1.]{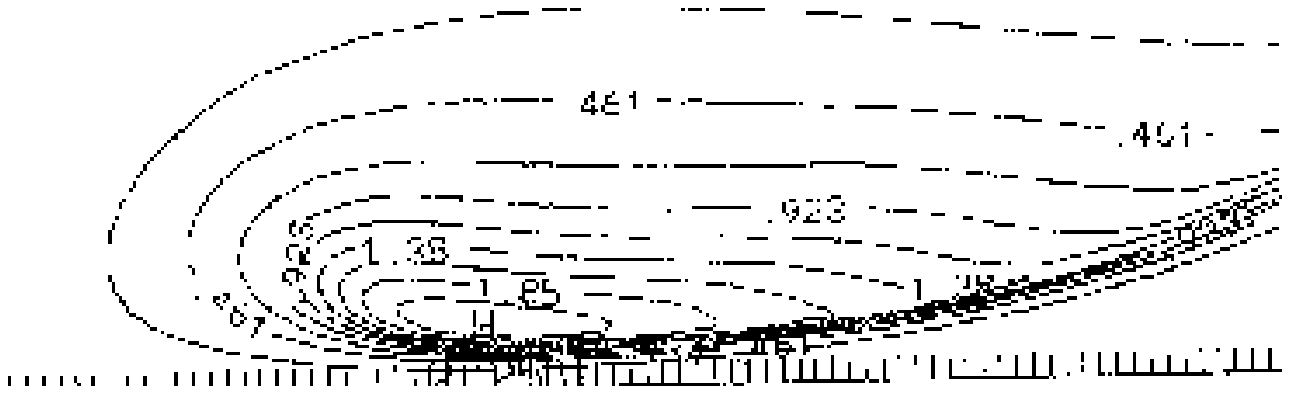}
\includegraphics[scale=1.5]{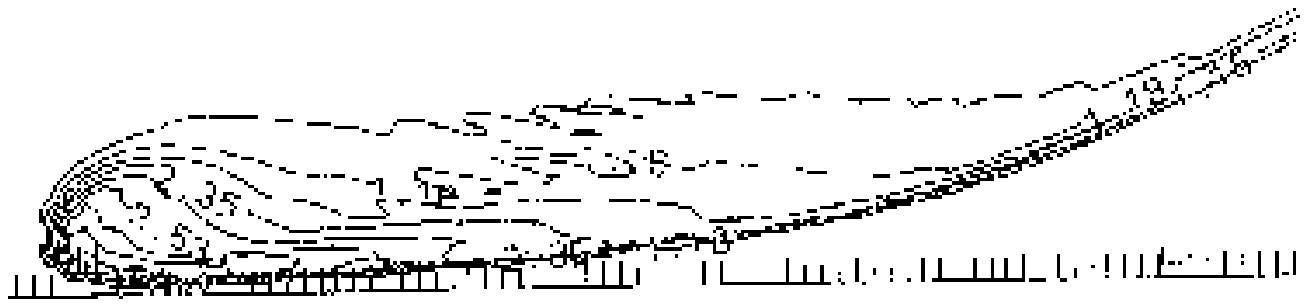}
\caption{$\omega$ in the symmetry plane, $z$ is not stretched, for
$t=0$, 6, 12 and 15.5} 
\label{fig:omp0}
\end{figure}

\section{Comparison of structures and \ominfty between older calculations}

This section will compare three calculations that meet the simulation
criteria (a-c) above:
\ITM \item \cite{Kerr93} 
\item An intermediate time from \cite{PumirSiggia90}.
\item An inviscid version of the \cite{ShelleyMO93} calculation.
\ITN

Earlier calculations such as \cite{AshurstMeiron87}, \cite{PumirKerr87}
and \cite{MelanderH89} met none of these criteria on coarse meshes.

When \cite{Kerr93} was published, the weight of numerical evidence from
simulations with similar initial conditions was against a singularity. 
Therefore, in making claims of singular behavior, 
besides demonstrating points (a-k) above,the following was
addressed:
\ITM\item The computational power of the calculations finding 
exponential growth was matched. 
\item A modification of the singular initial condition was found that was
able to reproduce the exponential behavior of competitive calculations.
\item A proposal was made for why there could be 
at least two types of behavior, exponential and singular.
\ITN

While \cite{PumirSiggia90} and \cite{Kerr93} meet all three 
requirements, the published work of \cite{ShelleyMO93} 
was neither inviscid nor did it have a localized perturbation.
A better description of \cite{ShelleyMO93} 
is that it was a slightly adaptive and higher Reynolds number version of 
\cite{MelanderH89}.  Nonetheless, by giving the profile they used
a localized perturbation and running it inviscidly, the
primary conclusion of \cite{ShelleyMO93}
could be verified. This was that for the 
\cite{MelanderH89} profile, singular behavior does not appear.  

The goal of working by steps through these three cases is to suggest how
the calculation of \cite{HouLi06} might be suppressing
singular behavior due to numerical noise in their numerical method. 
The first step follows \cite{Kerr92} to show how the initial condition 
of \cite{ShelleyMO93} leads to false regular behavior. 
Then the graphics for an inviscid version of \cite{ShelleyMO93} is compared to
the very late times of \cite{PumirSiggia90}. 
In this way the structures associated
with regular behavior will be identified. Finally, similar
structures are identified in \cite{HouLi06}.
This raises the possibility that the numerical method of \cite{HouLi06} 
also introduces small scale numerical errors.  
This will only be confirmed if their 2/3rds dealiasing calculations
are report or are repeated by someone else.

\begin{figure}
\includegraphics[scale=.75]{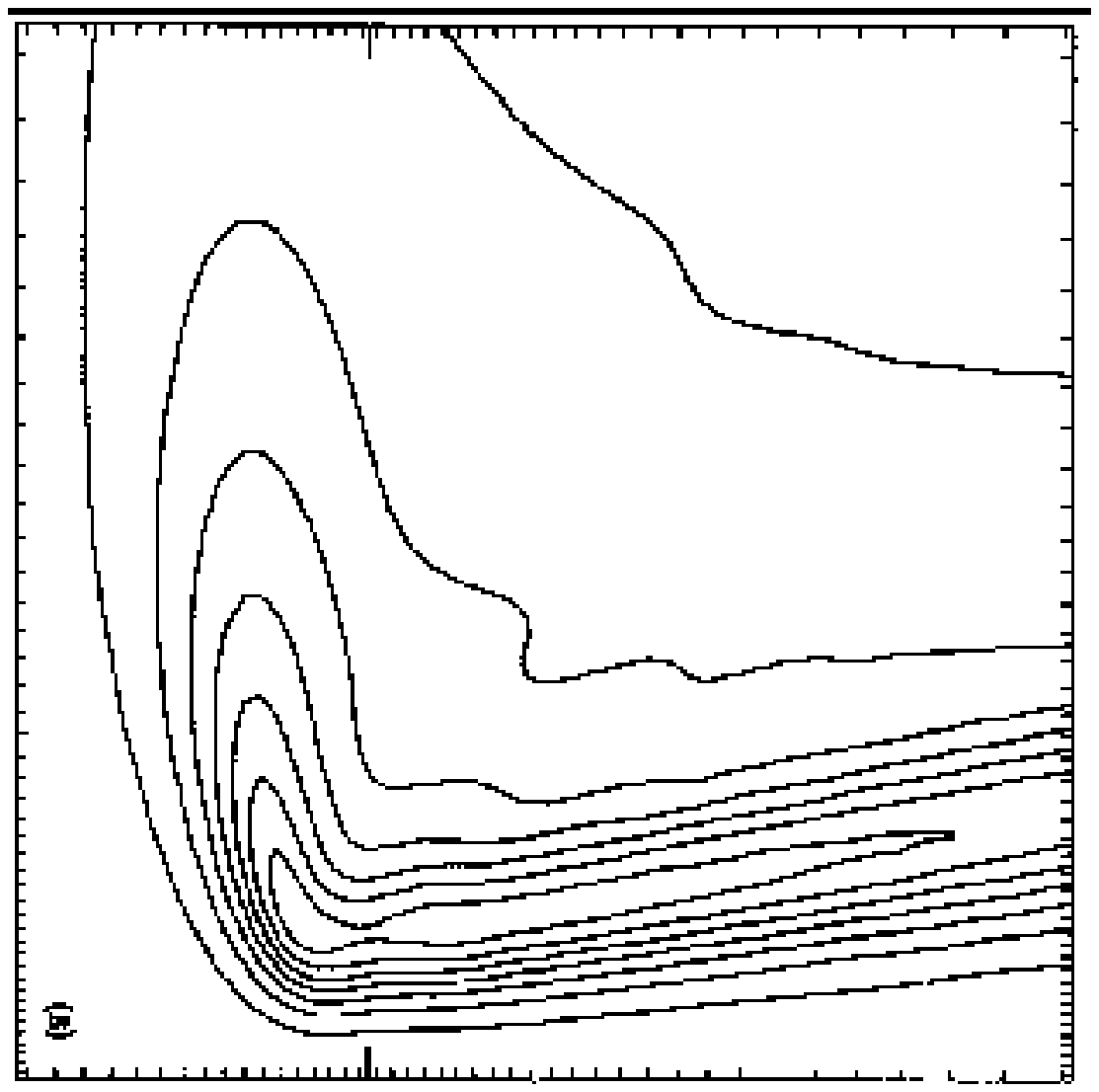}
\includegraphics[scale=.75]{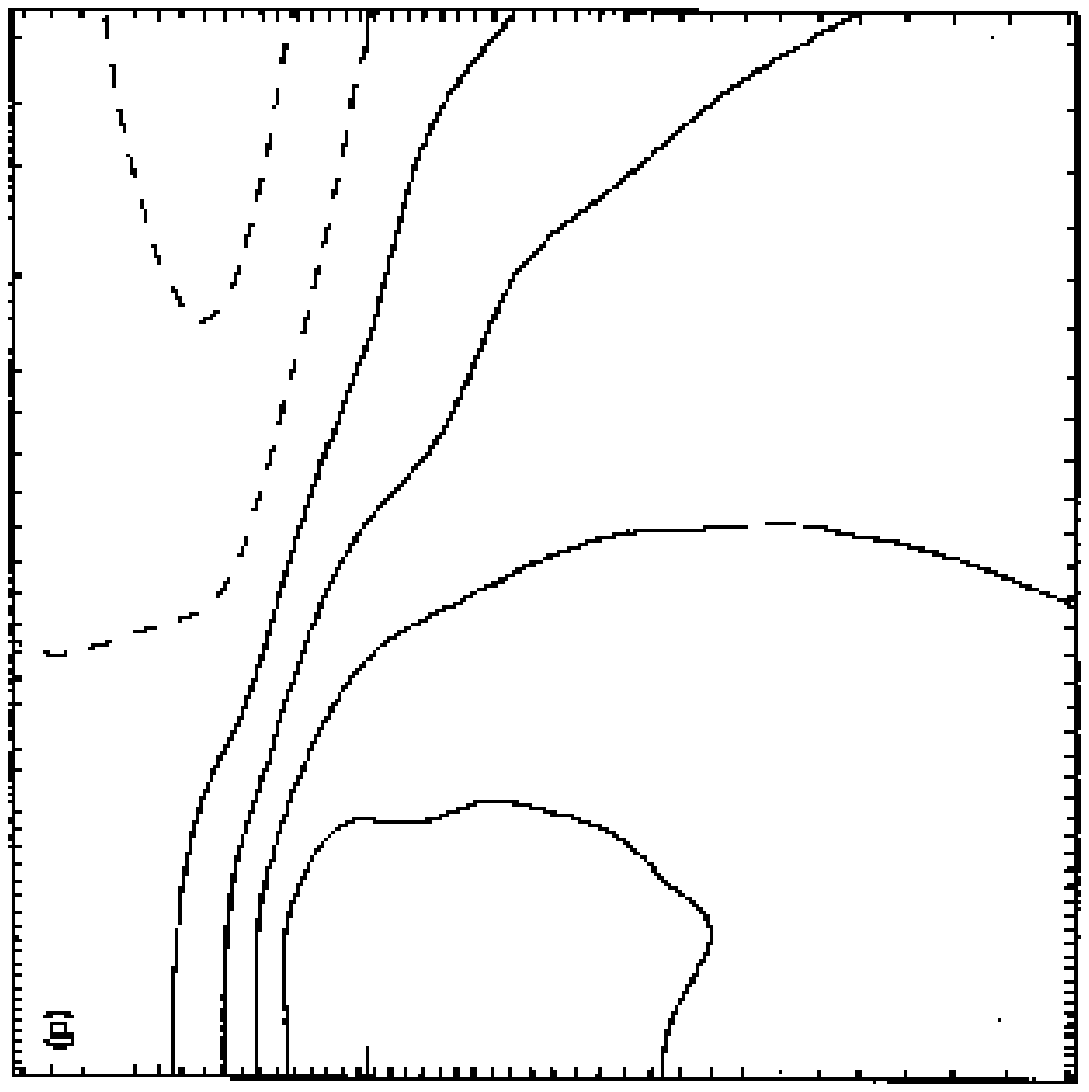}
\caption{$\ominfty$ and $\alpha$ contours in the symmetry plane for $t=7.2$ from
\cite{PumirSiggia90}} \label{fig:PSt7p2}
\includegraphics[scale=.9]{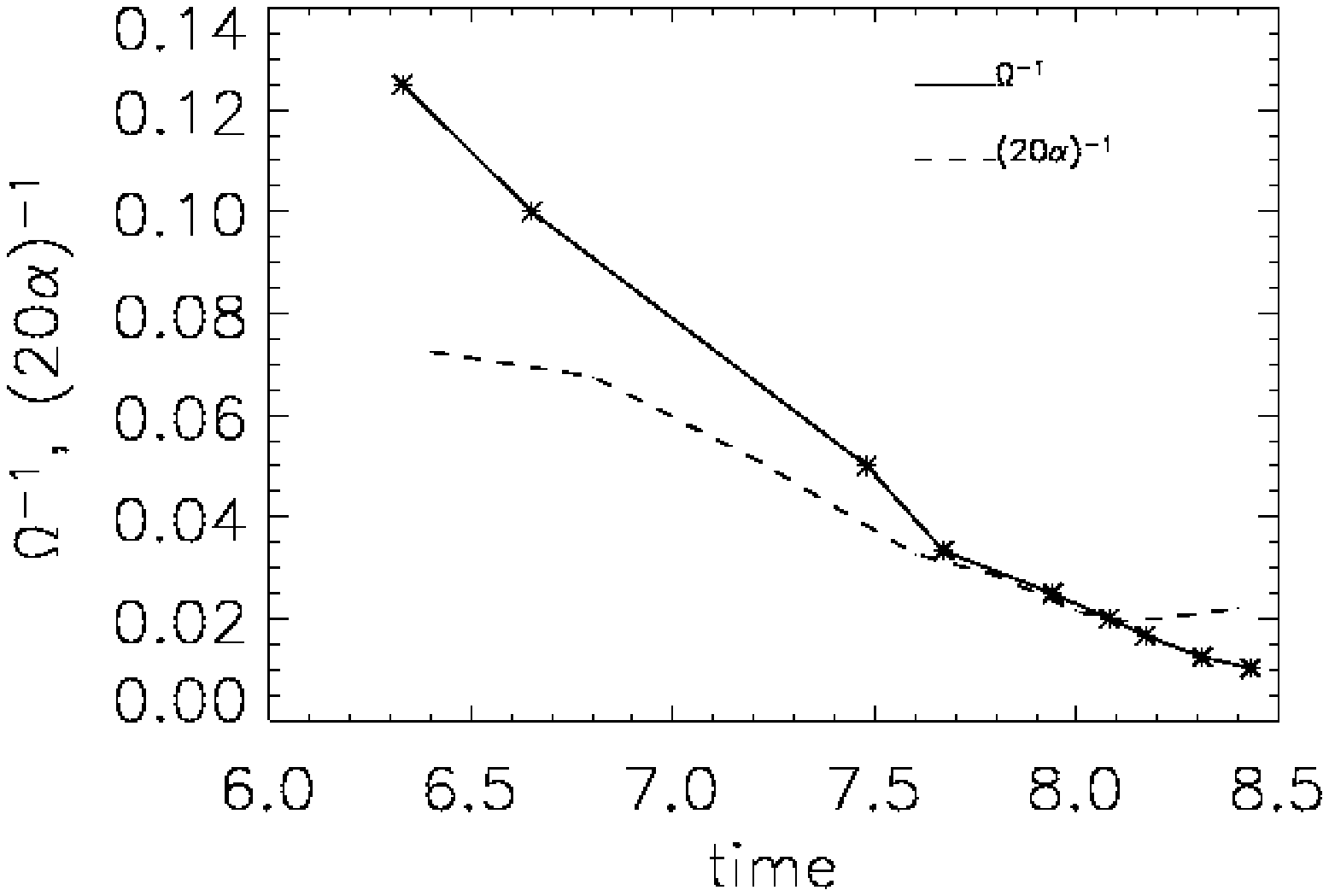}
\caption{Growth of $\ominfty$, $\alpha$ from
\cite{PumirSiggia90}} \label{fig:PS_oee}\end{figure}

\subsection{Comparing structures of \cite{Kerr93} and \cite{PumirSiggia90}}

The purpose of this subsection is to compare the structures that arise during
the period with the strongest evidence for singular growth of \cite{Kerr93} 
with a brief period of growth from
\cite{PumirSiggia90} that might have consistent behavior.  
Figure \ref{fig:omp} from \cite{Kerr05}
shows the growth of $\ominfty$, max($\alpha$)=$\|e_{yy}\|$ 
in the symmetry plane, and the enstrophy production $\Omega_{\rm pr}$.  
This was the primary evidence that these calculations have singular growth.  
Figure \ref{fig:omp1} plots the peak vorticity, peak strain along 
the vorticity in the symmetry plane, and the inverse of that strain 
as functions of time for a $256\times 256\times 192$ 
pseudospectral calculation using the initial condition 
of \cite{Kerr93}.  
This pseudospectral calculation was not the best
that was done in preparation for the final calculations 
of \cite{Kerr93}, but it can be used to illustrate the differences between
this initial condition and the \cite{MelanderH89} initial condition.
The peak of the $\alpha_\infty=\|e_{yy}\|_\infty$ strain is plotted 
instead of the value at the position of $\ominfty$ because it was 
better behaved as previously discussed (\cite{Kerr93}).

The vorticity and strain contours in Fig. \ref{fig:k93t15} are straight out 
of \cite{Kerr93} for $t=15$.  These plots
have been expanded by a factor of 5 in the vertical ($z$) direction. 
This time was chosen because it shows the least numerical noise while
being the regime showing singular growth of \ominfty.
Subsequent to 1993 it was realized that most of
the noise in the contours in \cite{Kerr93} comes from a high wavenumber 
spectral tail, 
which when filtered out gives the smoother figures of \cite{Kerr05}.  
It is likely that the physical space data for these fields in this plane 
is still stored on the NCAR mass store
and might be used later to create better figures.

The two points of comparison to be made are that:
\ENM\item The highest contour of vorticity is placed in the corner of the
bent front.
\item There is a steepening of the positive strain contours in the direction of 
propagation of the front.
\EEN 

To compare with competitive calculations whose graphic results
do not stretch the inter-vortical ($z$) direction, it is useful to show 
contour plots of vorticity in the symmetry plane without stretching
in Fig. \ref{fig:omp0}.  The evolution with time is as follows:
\ITM\item At $t=6$ there is only flattening.  
\item At $t=12$ a tilt in the innermost contour appears.  
\item This becomes more of a crescent in the bend 
at $t=15.5$, which is essentially an unstretched version of left frame of 
Fig. \ref{fig:k93t15} with the highest vorticity
contour only within the bend.
\ITN

How does this compare to \cite{PumirSiggia90}?  Recall that they
use hyperbolic trajectories, while all the other calculations use 
periodic trajectories. 
Therefore, definitive comparisons are out of the question. Nonetheless,
Fig. \ref{fig:PS_oee} shows that growth of $\ominfty$ 
and $\alpha_p$, the strain at the position of $\ominfty$, in
\cite{PumirSiggia90} is similar to Fig. \ref{fig:omp}.  
Recall that in \cite{Kerr93} 
it was found that while $\|e_{yy}\|_\infty\approx 0.1\ominfty$,
$\alpha_p\approx 0.05\ominfty$.  Therefore there is a short period 
between $t=7.6$ and $t=8.1$ where the growth of $\ominfty$ and $\alpha_p$ 
in \cite{PumirSiggia90} might be consistent with \cite{Kerr93}.  
Most of the growth of $\ominfty$ 
before $t=7.6$ is just to set up the structure.  
$t>8.1$, when saturation appears, is discussed in the next subsection.

Figure \ref{fig:PSt7p2} shows contours of $\ominfty$ and $\alpha$ 
just before this period at $t=7.2$.  As in Fig. \ref{fig:k93t15}, 
the highest contour of vorticity 
is placed in the corner of the bent front and there is a steepening of 
the positive strain contours in the direction of propagation of the front.  
There is a long tail in the vorticity plot, but it seems to be separating
from the contour in the corner. In these respects the figures 
are similar to \cite{Kerr93}.

The next subsection will show that when non-singular growth is observed, 
these properties are not observed.  

\subsection{Initial profile and why a localized perturbation with filtering
is used}

There are two critical components in forming the initial condition 
previously discussed by \cite{Kerr92} and \cite{Kerr93}.  
\ITM\item
First is using a vortex core profile that, before smoothing, is analytic as
originally proposed by \cite{MelanderH89}. 
As they claimed, their profile is far superior
to the cut-off Gaussian profile used by \cite{KerrH89}.
However, even here, based on spectra (see below), 
it was decided that this it was not smooth enough and 
a small high-wavenumber filter was necessary.
\item Second, there is the intertwined choices of the initial perturbation 
in the trajectory of the vortex tube and of the dimensions of the periodic 
domain in the axial direction.  This is discussed next.
\ITN

Following \cite{KerrH89} the trajectory of the center of the 
initial vortex tubes was defined as
$$ X(s) = {x_\circ} + {\delta_x}\cos(s) $$
$$ Z(s) = {z_\circ} + {\delta_z}\cos(s), \quad\rm{where} $$
$$ s(Y) = {y_2} + {L_y}{\delta_{y1}}\sin(\pi{y_2}/{L_y}) \quad\rm{and} $$
$$ {y_2} = Y + {L_y}{\delta_{y2}}\sin(\pi Y/{L_y}) $$
rather than a simple sinusoidal trajectory.

The incompressible vorticity vector for this trajectory was proportional to
$$\begin{array}{rcl} \omega_x & = & -\dfrac{\pi\delta_x}{L_y}
\left[1+\pi\delta_{y2}\cos\left(\dfrac{\pi y}{L_y}\right)\right]\times
\left[1+\pi\delta_{y1}\cos\left(\dfrac{\pi y_2}{L_y}\right)\right]
\sin\left(\dfrac{\pi s(y)}{L_y}\right) \\
\omega_y & = & 1 \\
\omega_z & = & -\dfrac{\pi\delta_z}{L_y}
\left[1+\pi\delta_{y2}\cos\left(\dfrac{\pi y}{L_y}\right)\right]\times
\left[1+\pi\delta_{y1}\cos\left(\dfrac{\pi y_2}{L_y}\right)\right]
\sin\left(\dfrac{\pi s(y)}{L_y}\right) \end{array} $$

Following a suggestion by Melander (private communication) the profile of
vorticity about this trajectory was given by
$$    \omega(r) = \exp(f(r)), $$
where
$$  f(r) = -\frac{r^2}{1-r^2} + r^2(1+r^2+r^4)   $$
and a distance $r$ from the center of the vortex core $(X,Y,Z)$ of
characteristic radius $R$ is
$$ r=|(x,y,z)-(X,Y,Z)|/R \quad\rm{for} \quad r\leq 1 $$

These formulae have been corrected based upon errors in the text of
\cite{Kerr93} noted by \cite{HouLi06}.  \cite{HouLi06} missed the
following additional misprint.  In \cite{Kerr93} and \cite{HouLi06} in
the formulae for $X(s)$ and $Z(s)$
(or their equivalents) , '$/L_x$' and '$/L_z$' respectively should both
be '$/L_y$'. 

For $\delta_{y1} = \delta_{y2} = 0 $ the trajectory is sinusoidal,
as used by \cite{MelanderH89}.  Two problems were then identified with 
the procedure of \cite{MelanderH89}, 
which were replicated by \cite{ShelleyMO93}.

The first problem identified by \cite{KerrH89}
was that if a sinusoidal perturbation is used then, 
as the two vortices collapse in the 
direction between themselves, they also expanded in the axial $y$ direction.  
When this expansion runs into its periodic image, growth in $\ominfty$ 
is suppressed.  Doubling the domain size in $y$ does not resolve
the problem. Therefore to localize the perturbation, \cite{KerrH89} and
\cite{Kerr93} chose the following values of $\delta$: 
$\delta_{y1} =0.5, \delta_{y2} =0.4, \delta_x =-1.6$and $\delta_z = 0$ and
$z_\circ = 1.57$ and $R=0.75$.

\begin{figure}
\includegraphics[scale=.7]{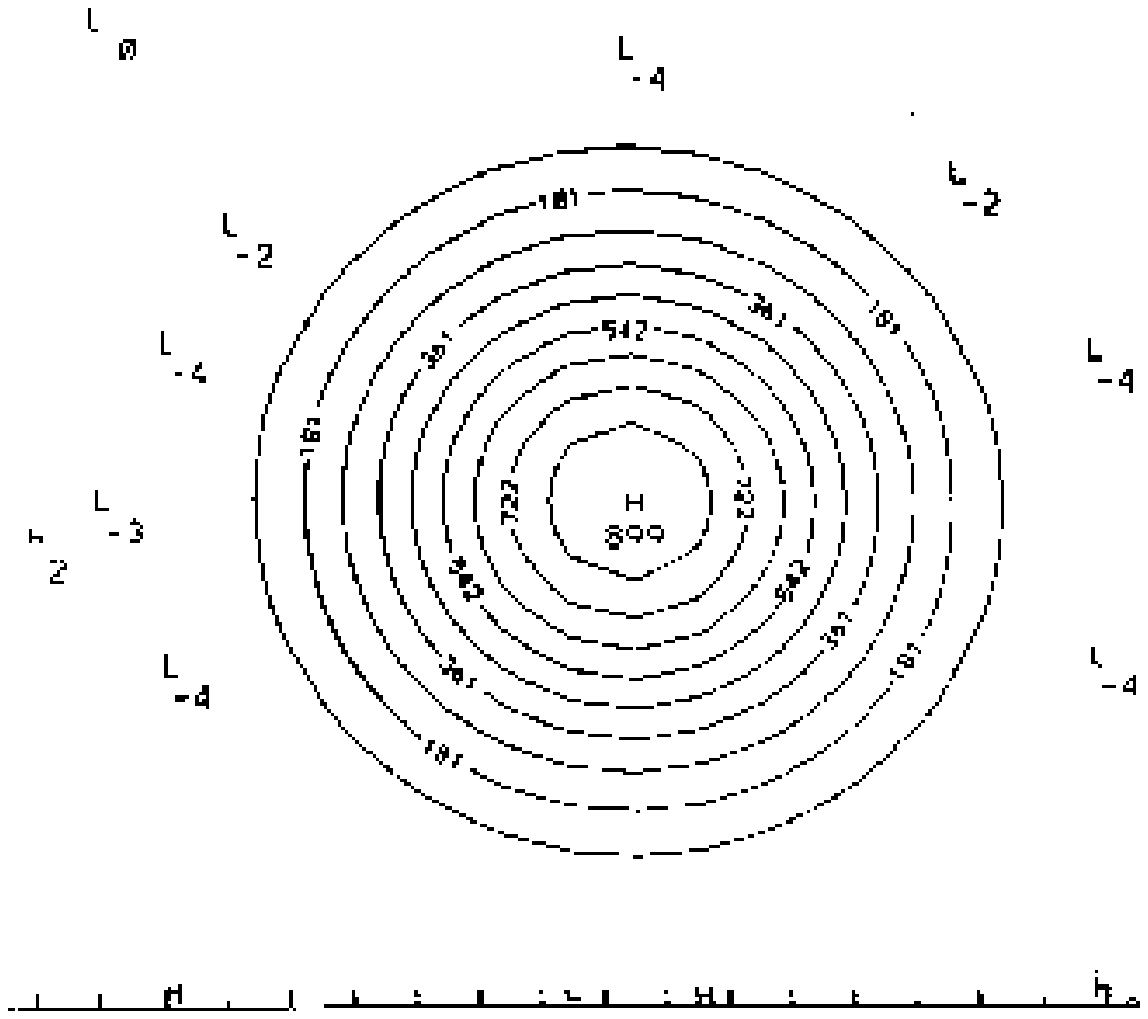}
\begin{minipage}[c]{.5 \textwidth}\vspace{-89mm}
\includegraphics[scale=.5]{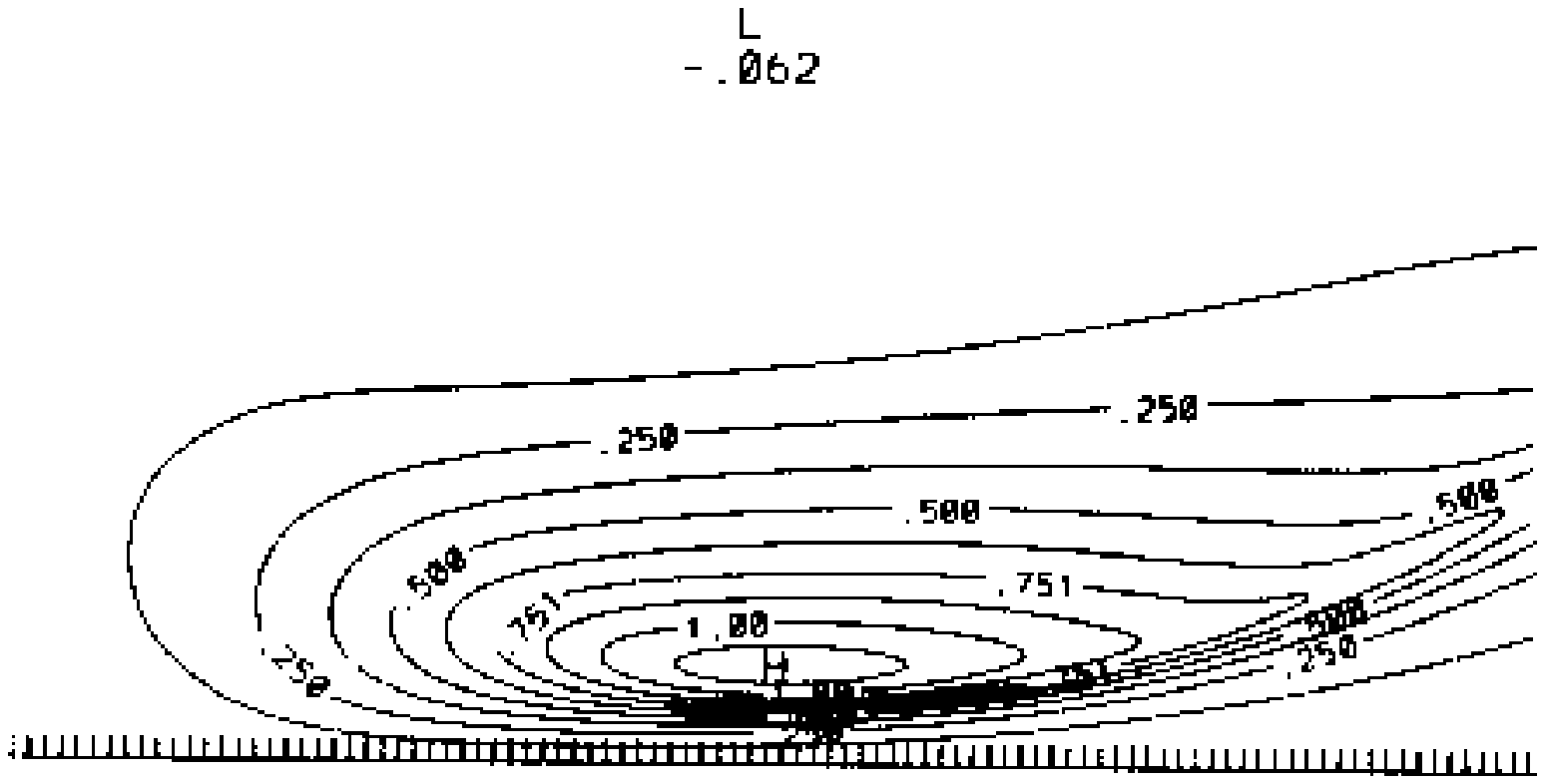}
\includegraphics[scale=.7]{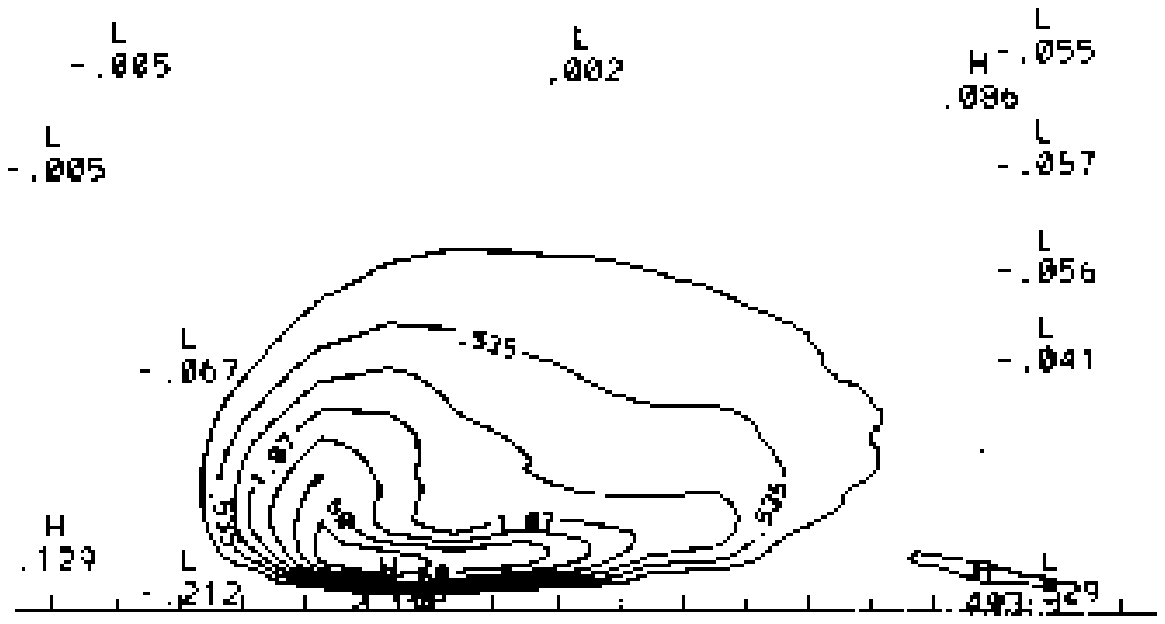}
\end{minipage}
  \caption[]{$y$ vorticity perpendicular in the symmetry plane.
$t=0$, $t=9$ and $t=13$ for \cite{Kerr93} calculation.}
  \label{fig:ITP1}
\end{figure}

A more serious problem was that the initial profile of 
\cite{MelanderH89}, without any filtering, created regions of 
negative vorticity in the symmetry plane.
There are small negative regions that cannot be seen
in the contour plots unless either zero contours are plotted or
highs and lows are given.  
This is demonstrated in the first frame of Fig. \ref{fig:ITP1}.

When highs and lows as in Fig \ref{fig:ITP1} are not used in the graphics, one
cannot assess the possible impact of small negative regions. 
This would apply to the graphics in \cite{PumirSiggia90},
\cite{ShelleyMO93} and now \cite{HouLi06} In \cite{HouLi06} negative contours
would not appear in the initial condition because their initial condition 
is filtered, but evidence for negative regions at late times is discussed 
in the next section.

\begin{figure}
\begin{minipage}[c]{.5 \textwidth}
\includegraphics[scale=.7]{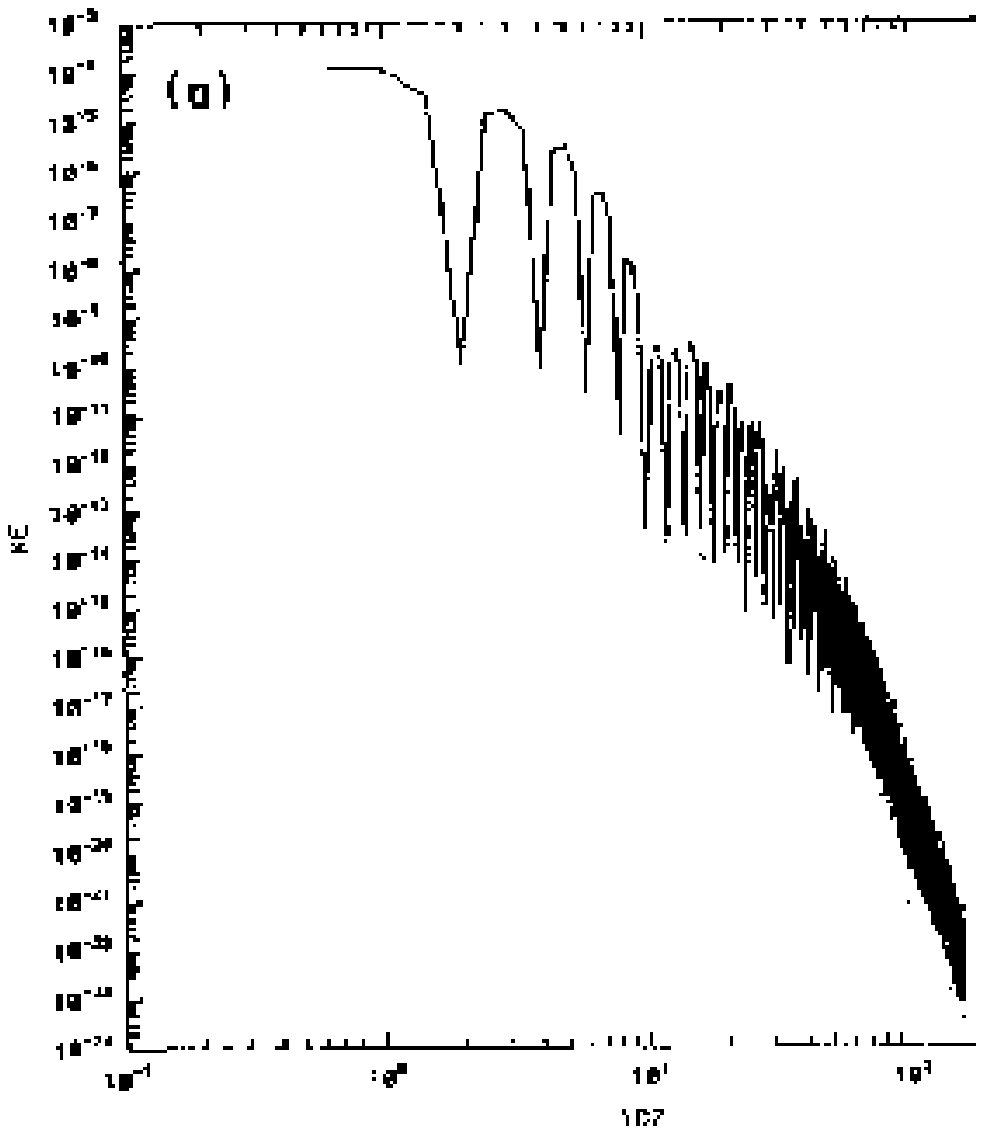}
  \caption[]{$z$-spectra for unfiltered initial conditions.}
  \label{fig:ITP2}
\end{minipage}
\hspace{5mm}
\begin{minipage}[c]{.4 \textwidth}\hspace{5mm}
\vspace{0mm}\hspace{-40mm}
\includegraphics[scale=.7]{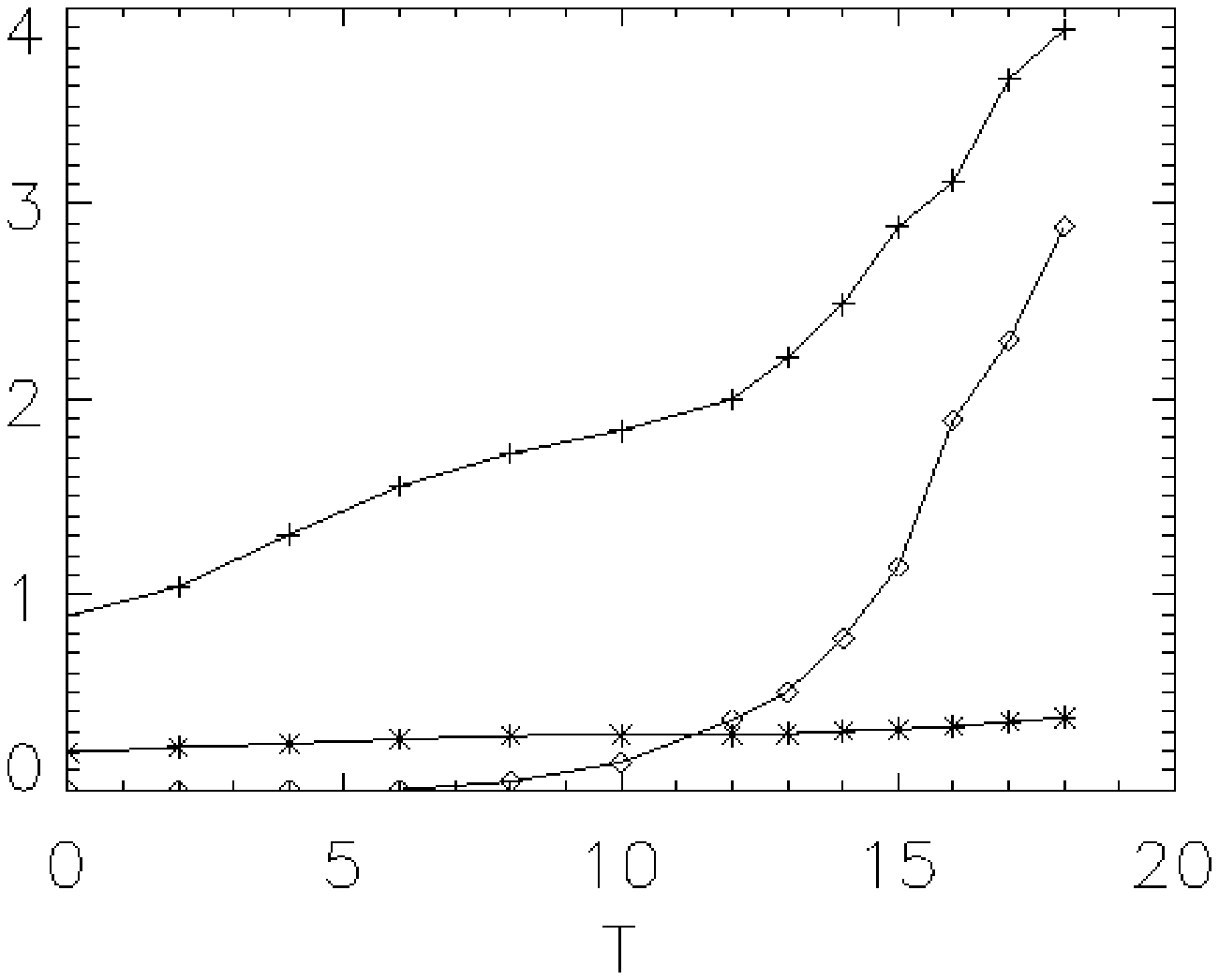}
  \caption[]{Peak strain along the vorticity (triangle), 
peak vorticity (circle), 
peak negative vorticity (plus) for unfiltered initial conditions vs. time.}
  \label{fig:ITP4}
\end{minipage}
\end{figure}

In preparation for the calculation in \cite{Kerr93} that used a filtered
initial condition, a variety of initial conditions without filtering
were experimented with.
A common feature of all the initial $z$ spectra, or equivalently 
the initial distribution of Chebyshev polynomials, was that they
went to zero with increasing wavenumber in a sawtooth power law
fashion.  An example spectrum is shown in Fig. \ref{fig:ITP2}. The envelope
of the high wavenumber sawtooth approximately obeys $k^{-2}$.  

If behavior where an exponential high wavenumber tail or
a power law $k^{-\gamma}$ with $\gamma>2$ is to be allowed,
this type of initial condition should not be used.  In fact,
\cite{Kerr93} found that $\gamma>3$ and associated $\gamma\rightarrow 3$
as the possible singular time was approached with
the test that $\Omega_{pr}\sim 1/(T-t)$ \ref{eq:ompr}.

Increasing the resolution used with the unfiltered initial conditions
did not alter either the low wavenumber components of the spectra or
the magnitude of the regions of negative vorticity.  That is, regions of
negative vorticity in roughly the same location and of the same magnitude 
appear.  This type of spectral behavior is reminiscent of Gibbs
phenomena, that is high wavenumber oscillations that occur when a
Galerkin method tries to represent physical space discontinuities.
This suggests that the origin of the sawtooth spectra and negative
regions of vorticity is the sharp cutoff in the vorticity profile
used by \cite{MelanderH89}, although possible inconsistencies
between the initial perturbation and periodic boundary conditions
might also have an influence.  This is despite the analytic formula
where all derivatives smoothly went to zero at the edge of the vortex
tube.  While this might work analytically, all that a numerical calculation
sees is that the vorticity goes from a finite value on one side of a
boundary to continuously zero on the other side.

\vspace{-2mm}
\subsection{Effect when there is no initial filter
\label{sec:Shelley}}

The calculation in this section is meant to show that if
the high wavenumber filter is not applied to the initial conditions,
then the strain saturates, afterwhich the
growth of the peak vorticity is exponential.  
Qualitatively everything about this calculation is the same as
the initialization used by \cite{Kerr93}. 
The only significant difference
was that the high wavenumber hyperviscous filter was not used.

Only enough resolution to demonstrate these points was used.  
There are some weak signs that at the last time calculated 
that the strain is beginning to increase and singular behavior 
might be starting.  However,
given the success of the filtered initial conditions there did
not seem to be any reason to follow this further.

The suggestion at the end of the last subsection is that isolated
points of negative vorticity might cause a problem.
The best evidence that these negative regions do affect the calculation is that
while they were initially very small, the order of $4\times 10^{-3}$ compared to
the initial peak vorticity of 1, by the end of the calculation
the negative peak was the same order of magnitude as the primary vortex.
This is shown in Fig. \ref{fig:ITP4}.

Figure \ref{fig:ITP4} also shows the strain, which saturates, leading to
exponential growth of peak vorticity.  It was also found that the position
of the peak of $\alpha$ is the same distance from the dividing
plane (in $z$) as the peak vorticity, whereas \cite{Kerr93} found
that the strain peak is always somewhat further from the dividing plane.

What are the differences in the structure?
The second frame at $t=9$ shows that the vorticity that develops from initial 
condition Fig. \ref{fig:ITP1} is not significantly different than 
the initial flattening from \cite{Kerr93}, 
except it is remaining flat longer.  
At the slightly later time of $t=13$, a head-tail structure is found 
as in earlier viscous calculations of vortex reconnection. The difference
with \cite{Kerr93} is that the highest contour is spread across a
vortex sheet above the lower dividing plane, across from its mirror image.

The spreading of this innermost contour is the best indication that similar 
behavior might be happening in other calculations.

How might these negative regions create this change in structure?
Detailed investigation of contours in preparing \cite{Kerr92} 
showed that
the largest negative regions had been sucked into the region between the
primary vortices, that is along the dividing plane.  They then grow
enormously and pair with their
opposite number across the lower dividing plane and induce velocity
opposite to that from the primary vorticity.  This pulls the tail out and
creates more flattening.

\begin{figure}
\includegraphics[scale=0.75]{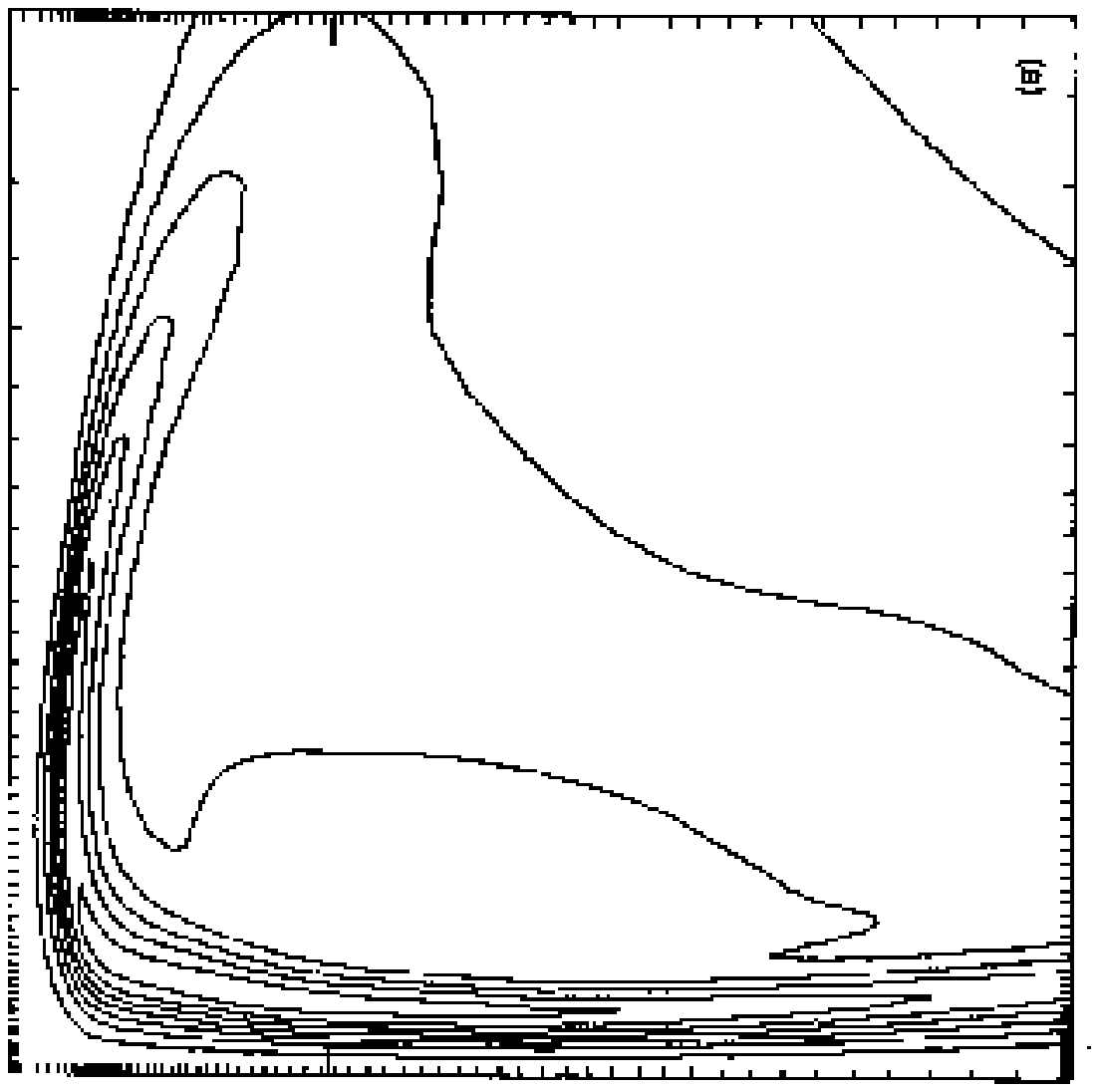}
\includegraphics[scale=0.75]{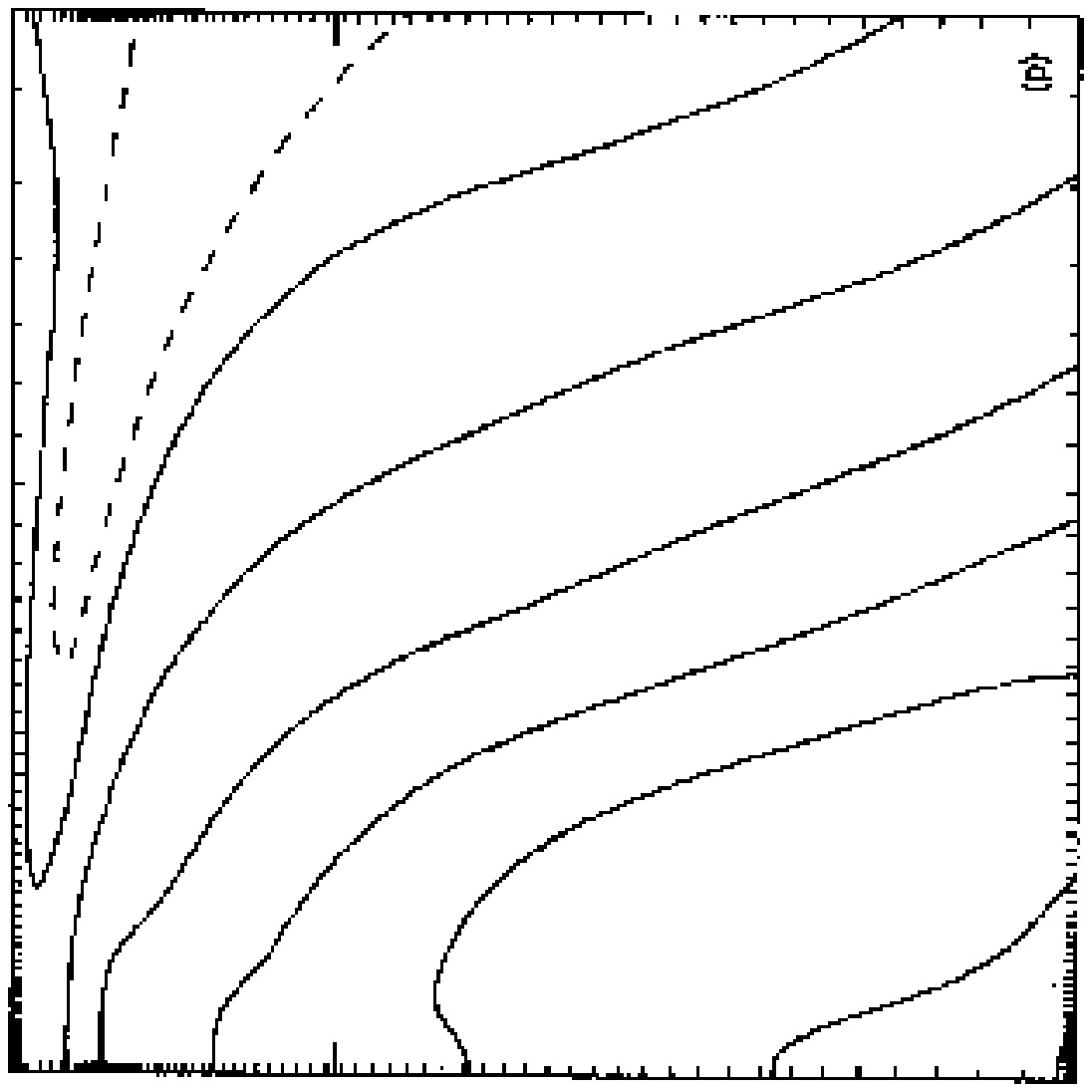}
\caption{$\ominfty$ and $\alpha$ contours in the symmetry plane for $t=8.3$ from
Pumir \& Siggia (1990).} \label{fig:PSt8p3}\end{figure} 
A comparison with the late times from \cite{PumirSiggia90} is now useful.  
That is for $t>8.1$, beyond when $\alpha_p\sim 0.05\ominfty$.
Recall that in Fig \ref{fig:PS_oee}, the strain was also observed to saturate
at late times, with exponential growth of the peak vorticity.  
Based on these comparisons, the late times of Pumir \& Siggia (1990)
should be checked to see if there are similar
regions of negative vorticity playing a major role,
and if so to determine whether these regions might then explain the saturation
of the strain in their calculations.
The highest vorticity contour is now spread over the entire tail
and the strain contours are not as concentrated.

\begin{figure}
\begin{minipage}[c]{.5 \textwidth}
\includegraphics[scale=.8]{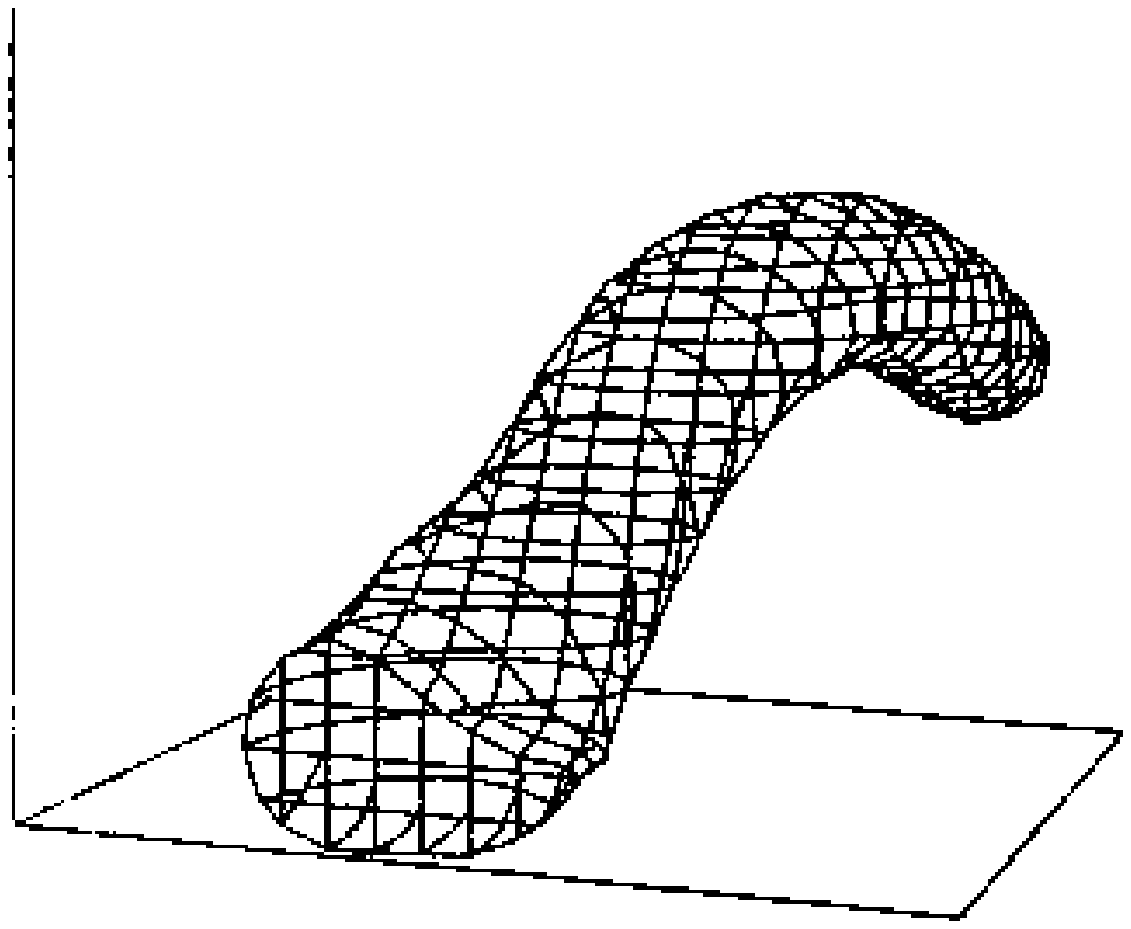}
  \caption[]{Three-dimensional isosurface plot of the vorticity
squared for $(x,y,z) = (4.9:8.8,0:3.9,0:3.9)$ at $t=12$ for unfiltered
initial conditions.  }
  \label{fig:ITP3}
\end{minipage}
\hspace{3mm}
\begin{minipage}[c]{.5 \textwidth}
\includegraphics[scale=.8]{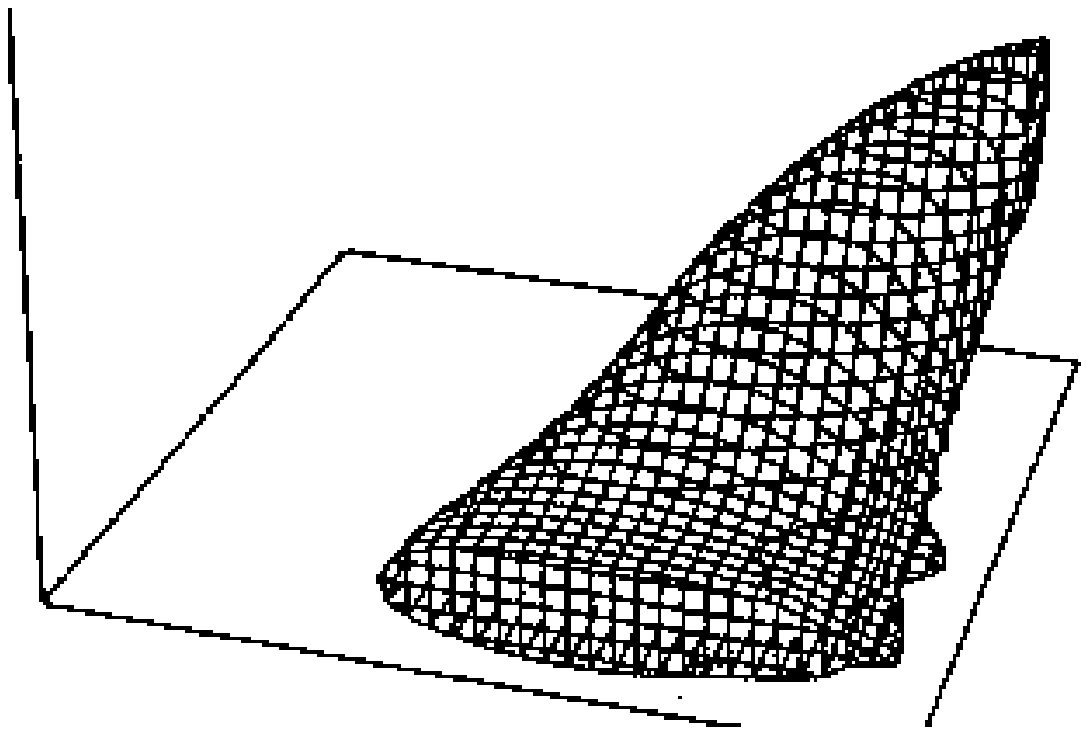}
\vspace{7mm}
  \caption[]{Three-dimensional isosurface plot of the vorticity
squared at $t=12$ for the filtered initial conditions 
in \cite{Kerr93}.  }
  \label{fig:Fig10t12}
\end{minipage}
\end{figure} 
Figure \ref{fig:ITP3} shows a three-dimensional isosurface plot of the vorticity
squared at $t=12$ for the unfiltered initial condition and
Fig. \ref{fig:Fig10t12} an isosurface for the filtered initial
condition at $t=12$.   Overall, the three-dimensional structure for
the unfiltered initial condition does not have 
two-dimensionalization of the vortices, one of the properties
observed by \cite{PumirSiggia90}, but does show less deformation 
than the filtered case from \cite{Kerr93}.
For the filtered initial condition, a crease in Fig. \ref{fig:Fig10t12} 
where the isosurface moves away from the dividing plane indicates 
where three-dimensional curvature of the vortex lines is occurring. A similar
crease does not appear for the unfiltered case.  

\begin{figure}
\includegraphics[scale=.9]{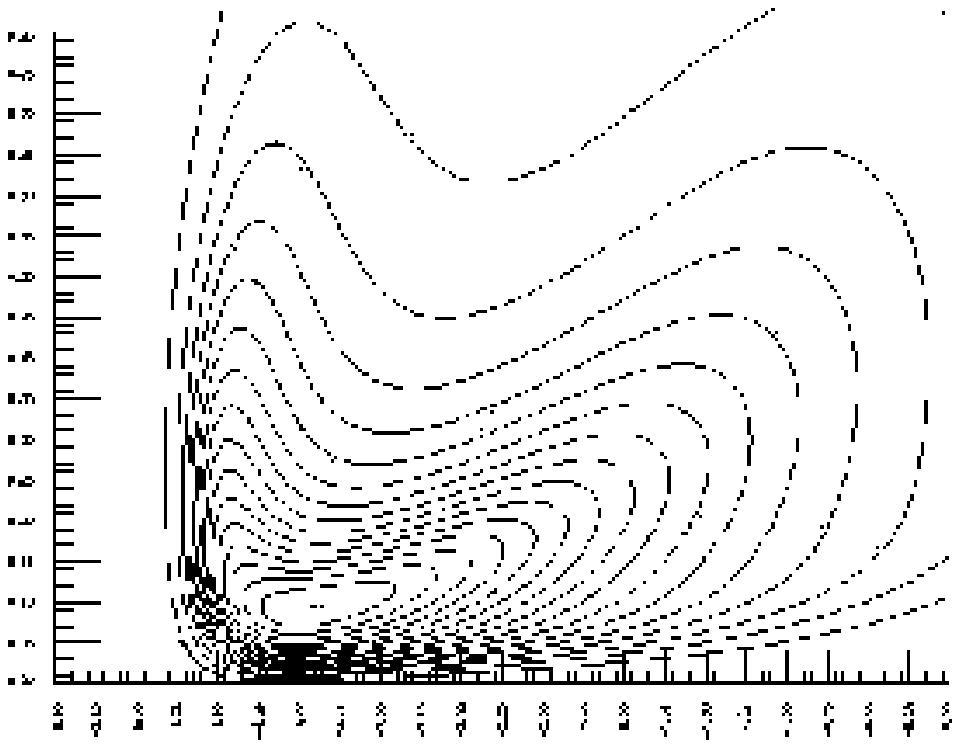}
\begin{minipage}[c]{.5 \textwidth}
\vspace{-55mm}
\includegraphics[scale=.6]{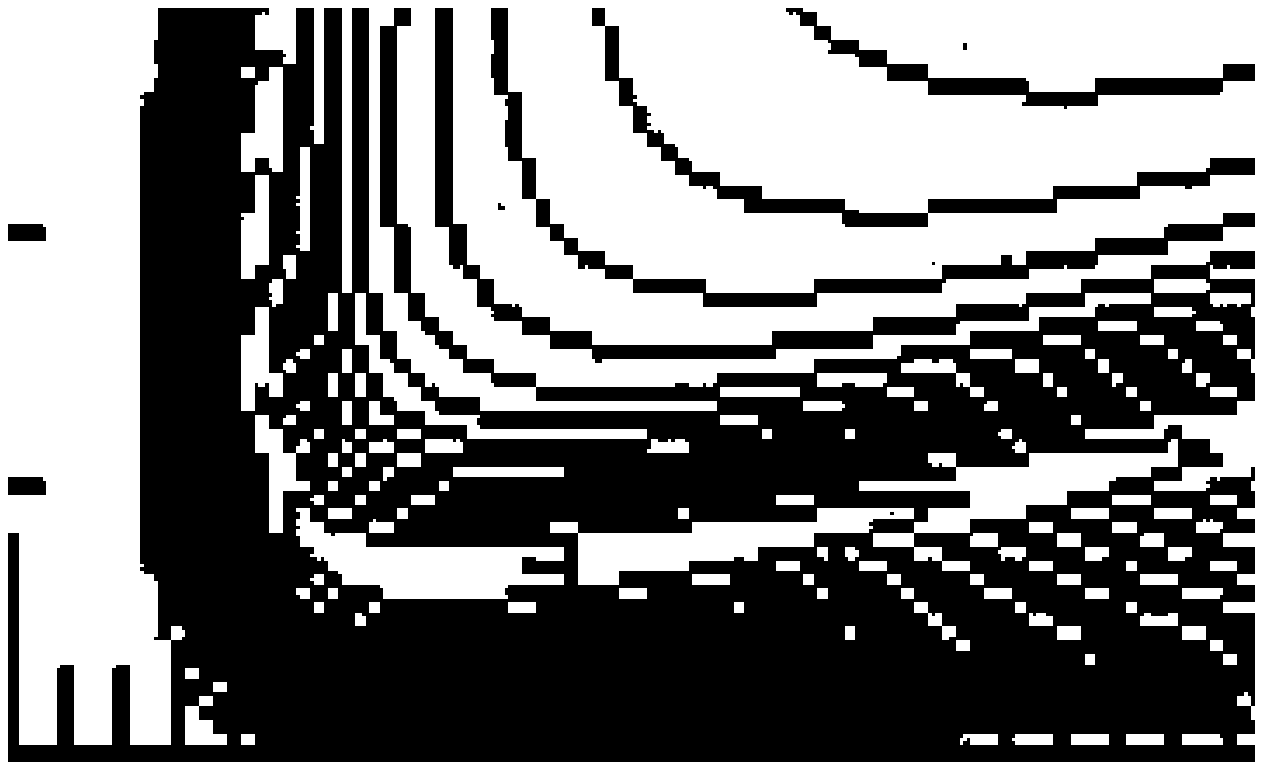}
\end{minipage}
  \caption[]{Contour plots through the symmetry plane from 
\cite{HouLi06} at $t=15$ and 17 (left and right) 
with the same factor of 
5 expansion in the vertical $z$ direction. The $t=17$ frame has been
blown up by a factor of 6.
Unlike Fig. \ref{fig:k93t15}, the $t=15$ contours here
look more like the $t=6$ contours in Fig \ref{fig:omp0}. Or $t=12$,
except there is no upward tilt on the left, even some downward tilt.
$t=17$ is very similar to the calculation with initial noise in 
Fig. \ref{fig:omp0}}
\label{fig:Hou1517} \end{figure} 

\begin{figure}
\includegraphics[scale=.57]{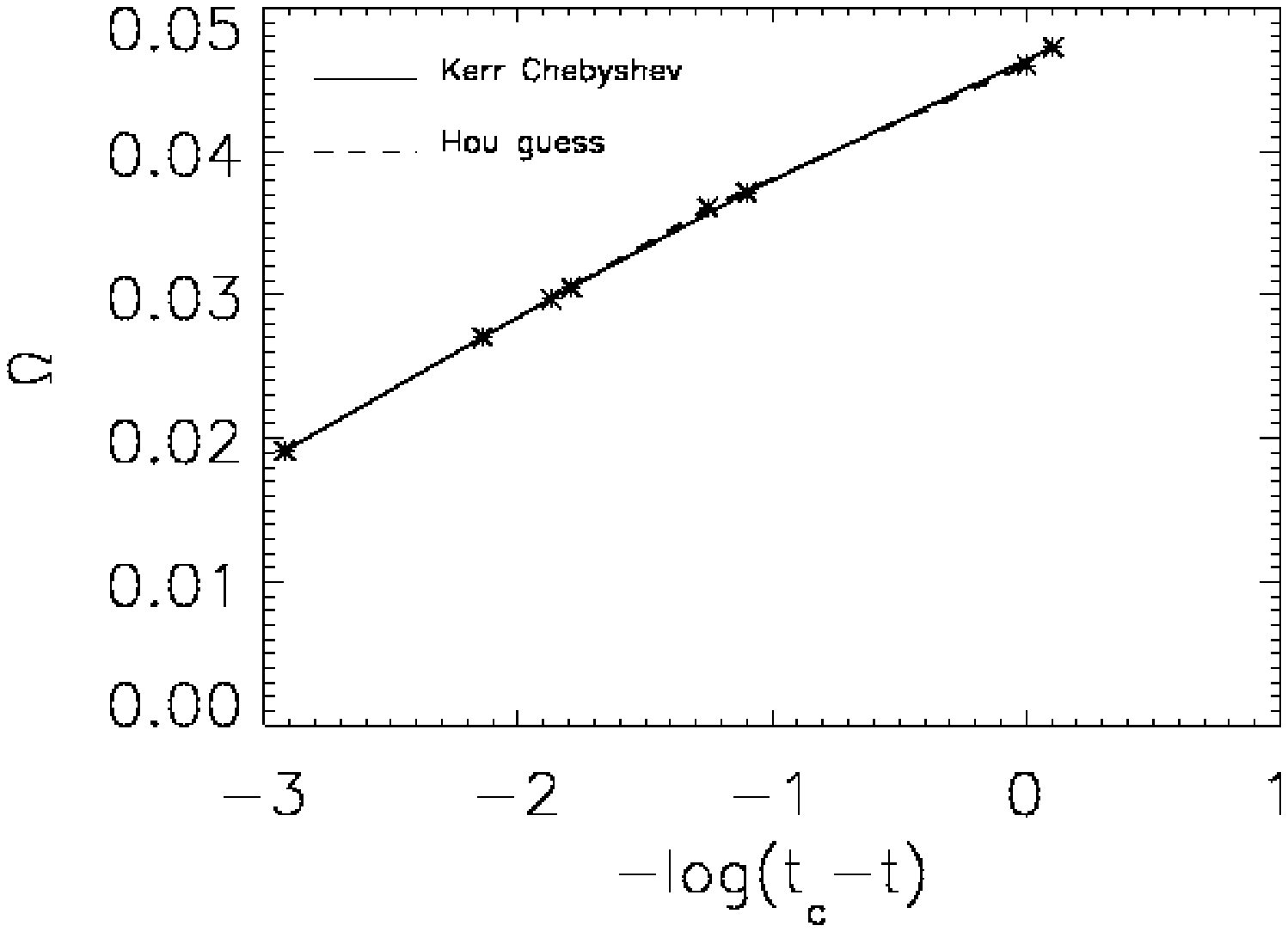}
\hspace{-8mm}\includegraphics[scale=.57]{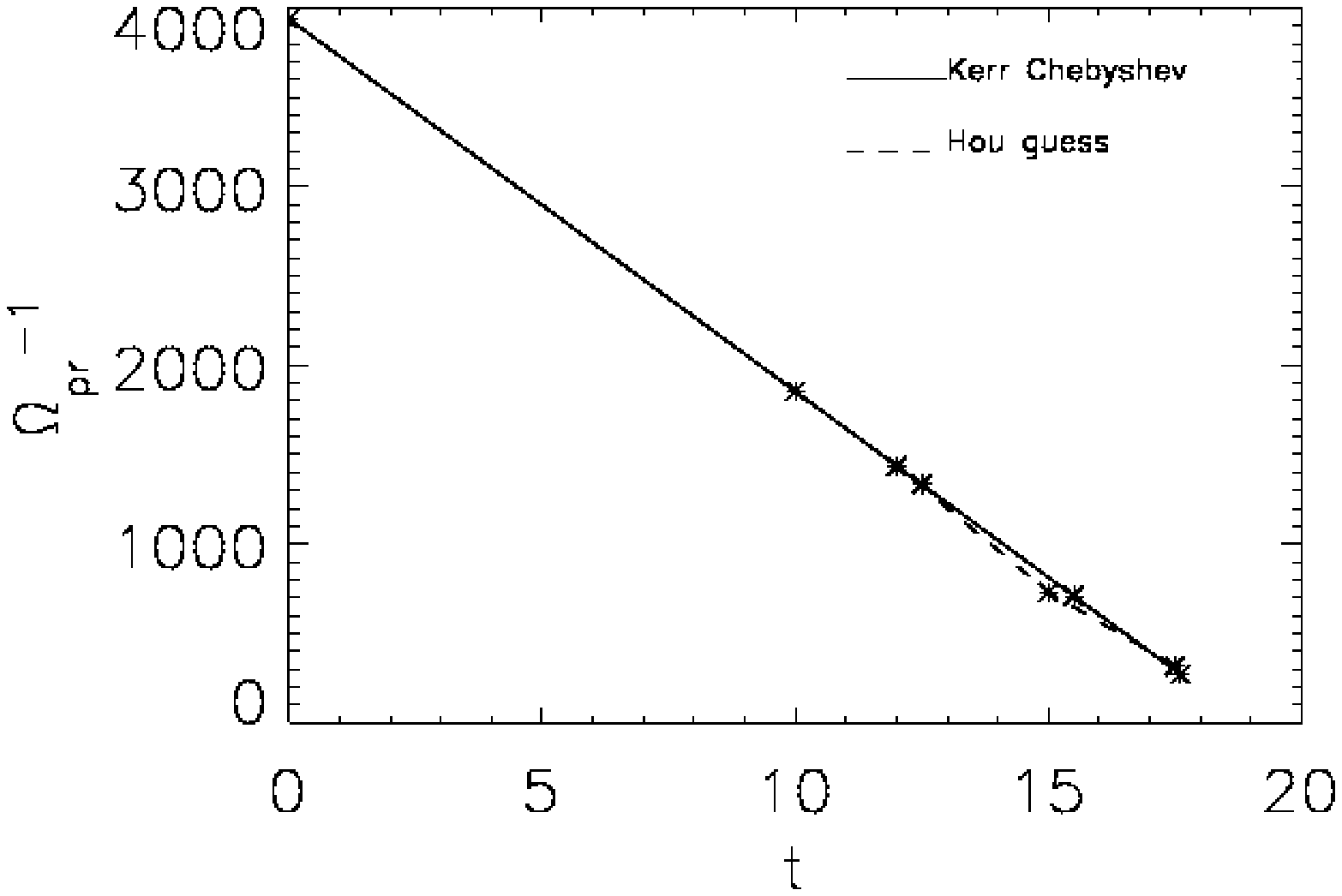}
\includegraphics[scale=.57]{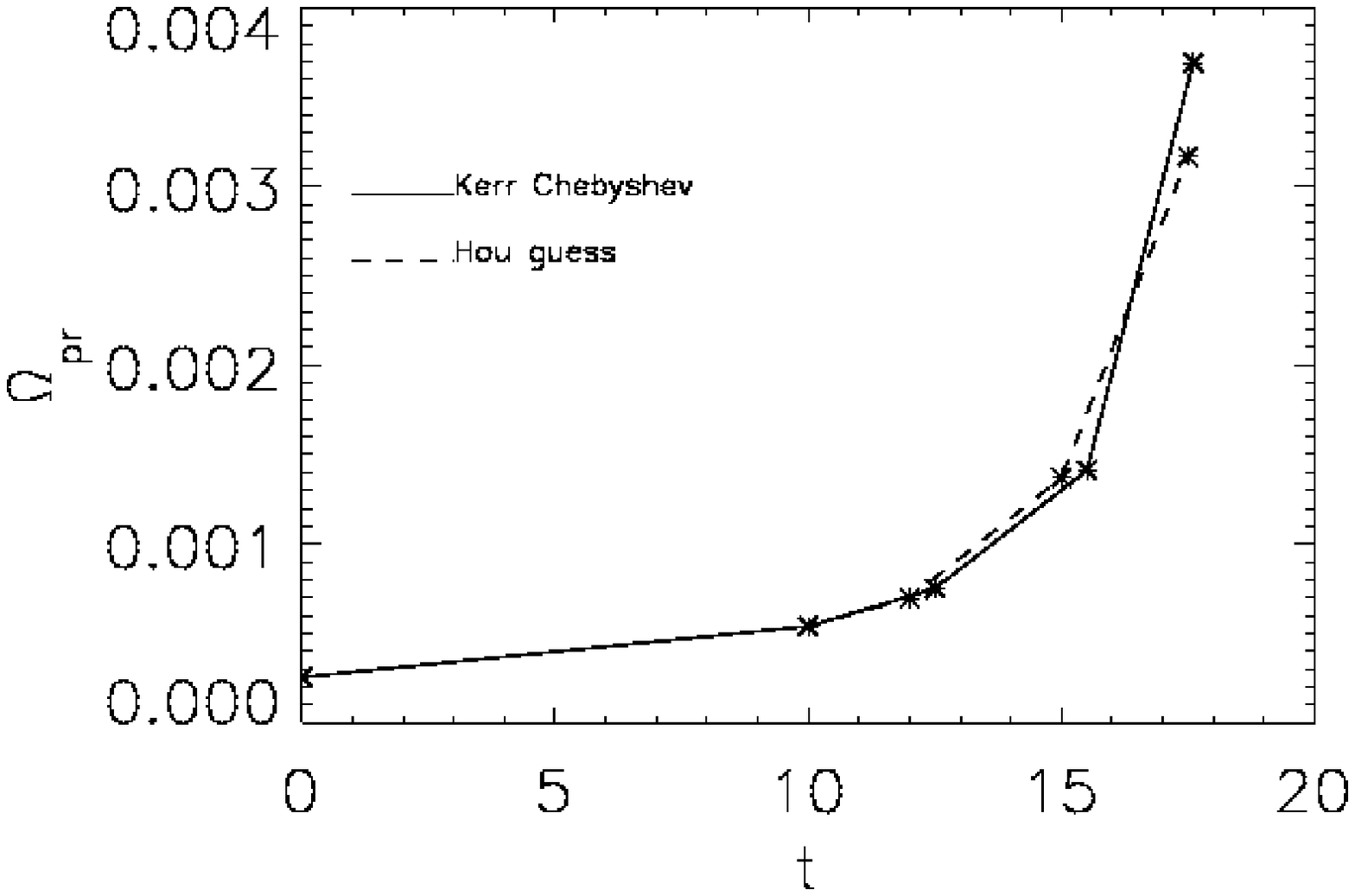}
\begin{minipage}[c]{.5 \textwidth}
\vspace{-70mm}
  \caption[]{Comparison of enstrophy and enstrophy production between
\cite{HouLi06} and \cite{Kerr93} 
assuming that they have the same
values at $t=10$.  The last two frames are $\Omega_{pr}$ not $\Omega$
as indicated.  In \cite{HouLi06} 
the growth of enstrophy is
greater for $12<t<17$, but is less for $t>17$.  The best signature
of this is the last frame of $\Omega_{pr}$.}
\end{minipage}
  \label{fig:HouLiom2}
\end{figure} 

\section{Discussion of new \cite{HouLi06} calculation \label{sec:HouLi}}

\cite{HouLi06} have recently released results where the initial
condition of \cite{Kerr93} was used exactly 
as originally programmed by using my old Fortran coding. 
That is: the same initial core profile, the same initial perturbation,
and the same high-wavenumber filtering.  As a result, exactly the
same initial behavior is expected and seems to be followed.

The numerical method was similar to that originally used by \cite{KerrH89}.  
That is pseudospectral using sine and
cosine transforms to apply the symmetries and different resolution
in the different directions.  The ratios of resolutions used was
different than in \cite{KerrH89}, but not significantly different.  
The highest resolution was $1536\times1024\times3072$.
3072 mesh points in the direction of greatest collapse makes the
local resolution at the dividing plane comparable to adaptive Chebyshev 
method of \cite{Kerr93}, while in the other directions the maximum
resolutions were respectively 3 and 4 times more than those used by the most
resolved case of \cite{Kerr93}.  
It was acknowledged by \cite{Kerr05}
that the resolution in $x$ was inadequate, although a simple post-processing
filter did allow new analysis in \cite{Kerr05}.

Another major difference was how the high wavenumbers were truncated.
This report will conclude that this is the most significant difference.
Both the hybrid pseudospectral/Chebyshev code reported in 
\cite{Kerr93} and the totally pseudospectral results from
around 1991 discussed here 
used the standard 2/3rds dealiasing, which turns a pseudospectral
calculation into a true spectral calculation.  
\cite{HouLi06} claim
that one can attain better convergence by allowing energy in
wavenumbers higher than the 2/3rds limit within a prescribed envelope.
The dangers of doing this are discussed here in section \ref{sec:uses}.

The comparison of the results of \cite{HouLi06} and \cite{Kerr93} can be
divided into the following three stages.
\ITM\item Early times, $t<10$,  where no results are reported by 
\cite{HouLi06} and for which this report will assume has 
identical behavior to \cite{Kerr93}.
\item Intermediate times $10<t<16.5$ where differences
are noted here but not by \cite{HouLi06}.
\item Late times where $t>16.5$ for which differences are reported by 
\cite{HouLi06} and where they claim that
they can successfully calculate until $t=19$.  
\ITN

Resolution and late times will be considered first so as to review 
the differences noted by \cite{HouLi06}.  Then the intermediate
stage will be discussed to show that differences appear much earlier.
This will be used to understand the late time differences.

\subsection{Resolution and late times}

Besides the calculation discussed in Sec. \ref{sec:Shelley} that used 
a fully pseudospectral calculation, among the tests done prior
to publication of the results in \cite{Kerr93} was
a fully pseudospectral calculation using nearly the same filtered, initial 
condition as \cite{Kerr93}.  The same initial peak vorticity and perturbation
were used. The only difference was a less severe initial filter.
The maximum resolution used was $256\times64\times512$ and the last
reliable time was estimated to be $t=15$ when $\ominfty=5.7$.  This calculation
is mentioned here only to note that tests were done to determine
what level of resolution would be needed to reach a given level
of singular behavior with a totally pseudospectral code.
My recollection that a remesh onto the $256\times64\times512$ mesh
from a $128\times64\times256$
was required at $t=13.5$ when $\ominfty=4$.  That is for every
desired doubling in $\ominfty$, the resolution needed to be quadrupoled
in at least the $x$ and $z$ directions.  Based upon this estimate and
assuming the singular growth reported by \cite{Kerr93},
with 3072 mesh points in $z$, the maximum $\ominfty$ that could be
obtained would be 14 at $t\approx 17.3$.

Based upon this, the claim by 
\cite{HouLi06} that they can calculate until $t=19$ would only
be possible if there isn't singular growth in $\ominfty$ and
the position of $\ominfty$ does not collapse as much as in \cite{Kerr93}.
As evidence that their calculation is resolved until that time, 
\cite{HouLi06} state that they still have 8 mesh points
between the position of $\ominfty$ and the dividing plane at that time.
A useful comparison between \cite{Kerr93} and \cite{HouLi06}
is Fig. 9 in \cite{HouLi06},
where there is a clear break in $\ominfty$ starting at $t=16.5$. 
from the $\ominfty\approx c/(18.7-t)$ result of \cite{Kerr93}.

From this one might conclude that the calculation of
\cite{Kerr93} should not have been run beyond about $t=16.5$.
However, the structural differences arise far before $t=16.5$. 

\subsection{Structural differences at $t=15$} 
Based on the comparisons of resolutions in \cite{Kerr93} 
and the pseudospectral test just mentioned, 
the evidence is that \cite{Kerr93} is resolved
at least until $t=16.5$.  Fig. \ref{fig:k93t15}, take from \cite{Kerr93} 
at $t=15<16.5$, shows that the innermost vorticity contour is in the corner.
In Fig. \ref{fig:Hou1517}, taken from Fig. 15 of \cite{HouLi06}, 
the innermost contour is flattened. 

What the contours from \cite{HouLi06} most resemble are the later times from
the unfiltered calculation from \cite{Kerr92} in Fig. \ref{fig:ITP1}
and $t>8.2$ from \cite{PumirSiggia90} 
shown in Fig. \ref{fig:PSt8p3}.  This suggests that noise
at the small scales that becomes amplified is the source of the differences.
Since they do filter their initial condition, where does the necessary
noise come from?  The most likely source would be retaining wavenumbers beyond
the 2/3rds rule cutoff.

The simplest test of this proposal would be to check for small negative regions
of vorticity, locate them, and plot how they grow in time. Our plan is
to run independent calculations to check this possibility. 

Another sign of a difference for $t<17$ is obtained by comparing the behavior of
enstrophy and its production plotted in Fig \ref{fig:HouLiom2}.  
Upon my suggestion, \cite{HouLi06} added the time dependence of enstrophy 
$\Omega$ and enstrophy production $\Omega_{pr}$
to the version they submitted. The units are not those used by 
\cite{Kerr93}, so
in Fig \ref{fig:HouLiom2} the assumption is made 
that the \cite{Kerr93} and \cite{HouLi06} calculations are identical 
for $t<10$ and therefore by 
scaling their values at $t=10$, a comparison can be made.  The discussion
in \cite{HouLi06} suggests that the values of 
$\Omega_{pr}$ shown were calculated from
$(d/dt)\Omega$ and not from $\int dV \omega_i e_{ij} \omega_j$ directly. 

Both $\Omega$ and $\Omega_{pr}$  grow faster than in \cite{Kerr93}
for $12<t<16$ and then more slowly.  This is seen best in the third figure
for $\Omega_{pr}$.  This trend
would be consistent with there being more enstrophy growth in the long tail
initially, but less concentration in the corner to build upon at later times.
In the table below the numbers used to make these plots are given.
The enstrophy results from \cite{Kerr93} correct a misprint.
The formula for enstrophy in \cite{Kerr93}
should have been: $\Omega=-.0105\log(18.9-t)+.05$, 
giving $\Omega(t=0)=.019$.

\BEA{ccccc} & {\rm Chebyshev} & t_c=18.9 &  
{\rm Hou \& Li} & t_c=18.7\\
{\rm time} & 15.5 & 17.5 & 15 & 17.5 \\
\Omega & .0375 & .0507 & 6.9 & 9.8 \\
t_c-t & 3.4 & 1.4 & 3.7 & 1.2 \\
\log(t_c-t) & 1.22 & .336 \\
\Omega_{pr} & .00158 & .00375 & .6 & 1.95 \EEA

\section{Moving forward}

In March 2003, A. Bhattacharee, U. Frisch, R.M. Kerr, N. Zabusky and others met 
at the Institute for Advanced Studies in Princeton to consider 
what direction a new computational effort should follow.  This meeting was 
instigated by the untimely death of our friend and collegue Rich Pelz. 

The primary conclusion that was reached was that there was a serious problem 
where each numerical method seemed to give a different answer.  
It was concluded that this was primarily because each team
was pushing their calculations a little too far.  In the early 1990s 
there was really no choice. 
To get any trend one had to overextend the calculations.  
In the first decade of the 21st century there is the luxury of 
far greater computing power and significantly improved numerical methods.  
Therefore it was decided to propose
an international collaborative effort based on:
\ITM \item A factor of at least 4 increase in resolution in each direction
\item Availability of modern adaptive mesh finite difference and spectral element
codes capable of providing significant improvements in local resolution.
\item Separate groups should conduct simultaneous simulations with the same
initial conditions.  The only results that should be reported are until 
a time when the two calculations differ only by an agreed upon amount.  
\ITN

Only in the UK was a proposal made.  A separate proposal with no 
communication with the people above, and therefore not informed of the 
rules, was funded and went to T. Hou at Cal Tech, R. Caflisch at UCLA
and M. Siegel at the New Jersey Institute of Technology.  The UK
proposal between Warwick and Imperial was never funded due to the poor
rules for reviewing interdisciplinary proposals in the UK.  
Funding has finally been obtained from the Leverhulme Foundation,
but only for one post-doc at Warwick.  

The original intention was that the two first adaptive codes to be used 
would be the spectral element code of Barkley at Warwick and 
some variant of the finite-difference adaptive code of Grauer in Bochum 
or from within Bhattacharjeee's group at New Hampshire.  Later, 
the improved spectral element code of Sherwin from Imperial would be used.

Most likely at this time we will go directly to Sherwin's spectral element code 
and work with Pumir on reviving his adaptive-mesh finite difference code.

However, the first priority now is to address the issues raised by 
\cite{HouLi06}.  
Therefore, their calculations will be repeated and
compared in detail with a straight pseudospectral calculation using
a strict 2/3rds rule and stopped early enough so that there are no
questions about small-scale resolution.  This should be near $t=17$,
although the comparison above suggests that a difference should
show up as early as $t=15$.


\begin{thebibliography}{}
\bibitem[Ashurst \biband Meiron (1987)]{AshurstMeiron87} %
\authtwo{W.}{Ashurst}{D.}{Meiron}\yprl{1987}
{58}{1632}{-1635}{Numerical study of vortex reconnection}

\bibitem[Beale \etall (1984)]{BKM84} %
\auththr{J. T.}{Beale}{T.}{Kato}{A.}{Majda}
\yjour{1984}{Commun. Math. Phys.}{94}{61}{}
{Remarks on the breakdown of smooth solutions for the $3D$ Euler equations}

\bibitem[Boratav \etall (1992)]{BPZ92} %
\auththr{ON}{Boratav}{RB}{Pelz}{NJ}{Zabusky}
\ypfa{1992}{4}{581}{-605}
{Reconnection in orthogonally interacting vortex tubes: Direct numerical
simulations and quantification in orthogonally interacting vortices}

\bibitem[Brachet \etall (1983)]{Brachetetal83} %
\authmanythr{M.E.}{Brachet}{D.I.}{Meiron}{S. A.}{Orszag}\auththr{B.
G.}{Nickel}
{R.H.}{Morf}{U.}{Frisch}\yjfm{1983}{130}{411}{452}{Small-scale structure
of the Taylor-Green vortex}

\bibitem[Caffarelle \etall (1982)]{Caffarelleetal82}
\auththr{L.}{Caffarelle}{R.}{Kohn}{L.}{Nirenberg}
\yjour{1982}{Commun. Pure Appl.  Math.}{35}{771}{}{}

\bibitem[Constantin \etall (1996)]{ConstFM96} %
\auththr{P.}{Constantin}{C.}{Fefferman}{A.}{Majda}
\yjour{1996}{Comm. Partial. Diff. Equns.}{21}{559}{-571}
{Geometric constraints on potentially singular solutions for the 
3D Euler equations}

\bibitem[Grauer \etall (1998)]{Graueretal98} %
\auththr{R.}{Grauer}{C.}{Marliani}{K.}{Germaschewski} 
\yprl{1998}{80}{4177}{-4180}{Adaptive mesh refinement for singular solutions of 
the incompressible Euler equations.} 

\bibitem[Hou \biband Li (2006)]{HouLi06} \authtwo{T.Y.}{Hou}{R.}{Li}
\sjour{2006}{Dynamic depletion of vortex stretching and
non-blowup of the 3-D incompressible Euler equations}
{J. Nonlin. Sci.}

\bibitem[Kerr (1992)]{Kerr92}\authone{R.M.}{Kerr} 
\yproc{1992}{309}{-336}{Evidence for a singularity of the three-dimensional
incompressible Euler equations.}
{Topological aspects of the dynamics of fluids and plasmas}
{G.M. Zaslavsky, M.  Tabor \biband P. Comte}
{Proceedings of the NATO-ARW workshop at the Institute for Theoretical Physics, 
University of California at Santa Barbara. 
Kluwer Academic Publishers, Dordrecht, Netherlands.}

\bibitem[Kerr (1993)]{Kerr93} %
\authone{R.M.}{Kerr} \ypfa{1993a}{5}{1725}{-1746}{Evidence for a singularity
of the three-dimensional, incompressible Euler equations}

\bibitem[Kerr (2005)]{Kerr05}\authone{R.M.}{Kerr}\ypf{2005}{17}{075103}{}
{Velocity and scaling of collapsing Euler vortices}

\bibitem[Kerr \biband Hussain (1989)]{KerrH89} %
\authtwo{R.M.}{Kerr}{F.}{Hussain}\yjour{1989}
{Physica D}{37}{474}{-484}{Simulation of vortex reconnection}

\bibitem[Melander \biband Hussain (1989)]{MelanderH89} %
\authtwo{M.V.}{Melander}{F.}{Hussain}\ypfa{1989}{1}{633}{636}
{Cross-linking of two antiparallel vortex tubes}

\bibitem[Pelz (2001)]{Pelz01}\authone{R.}{Pelz}
\yjour{2001}{J. Fluid Mech.}{444}{299}{-320}
{ Symmetry and the hydrodynamic blow-up problem}

\bibitem[Ponce (1985)]{Ponce85}\authone{G.}{Ponce} %
\yjour{1985}{Commun. Math. Phys.}{98}{349}{}
{Remark on a paper by J.T. Beale, T. Kato and A. Majda}

\bibitem[Pumir \biband Kerr (1987)]{PumirKerr87} %
\authtwo{A.}{Pumir}{R. M.}{Kerr}\yprl{1987}{58}{1636}{-1639}
{Numerical simulation of interacting vortex tubes}

\bibitem[Pumir \biband Siggia (1987)]{PumirSiggia87} %
\authtwo{A.}{Pumir}{E. D.}{Siggia}\ypf{1987}{30}{1606}{-1626}
{Vortex dynamics and the existence of solutions of the Navier-Stokes equations}

\bibitem[Pumir \biband Siggia (1990)]{PumirSiggia90} %
\authtwo{A.}{Pumir}{E. D.}{Siggia}\ypfa{1990}{2}{220}{-241}
{Collapsing solutions to the 3-D Euler equations}

\bibitem[Shelley et al. (1993)]{ShelleyMO93} %
\auththr{M.J.}{Shelley}{D.I.}{Meiron}{S.A.}{Orszag}
\yjfm{1993}{}{246}{613}{}

\bibitem[Sulem et al. (1985)]{Sulemetal85} %
\authfour{P.L.}{Sulem}{U.}{Frisch}{A.}{Pouquet}{M.}{Meneguzzi}
\yjour{1985}{J. Plasma Phys.}{33}{191}{}{}

\bibitem[Kerr (2006)]{Kerr06}
\authone{R.M.}{Kerr} 2006 Computational Euler History. 
Identical text with cleaner figures are in:\hfil\break
http://www.eng.warwick.ac.uk/staff/rmk/rmk\_pubs/compEulerhist\_arXiv.pdf


























\end{thebibliography}
\end{document}